\documentclass[aps,prb,reprint,showpacs,floatfix]{revtex4-1}
\usepackage{graphicx}  % needed for figures
\usepackage{dcolumn}   % needed for some tables
\usepackage{bm}        % for math
\usepackage{amssymb}   % for math
\usepackage{amsmath, amsthm}
\usepackage[extra]{tipa}
\usepackage{color}
\setcitestyle{square,numbers}

\begin{document}

\title{Correlated atomic wires on substrates. I. Mapping to quasi-one-dimensional models}
\author{Anas Abdelwahab}
\author{Eric Jeckelmann}
\affiliation{Leibniz Universit\"{a}t Hannover, Institut f\"{u}r Theoretische Physik, Appelstr.~2, 30167 Hannover, Germany}
\author{Martin Hohenadler}
\affiliation{\mbox{Institut f\"ur Theoretische Physik und Astrophysik,
    Universit\"at W\"urzburg, Am Hubland, 97074 W\"urzburg, Germany}}

\date{\today}

\begin{abstract}
We present a theoretical study of correlated atomic wires deposited on
substrates in two parts. In this first part, 
we propose lattice models for a one-dimensional quantum wire on a
three-dimensional substrate and map them onto effective two-dimensional
lattices using the Lanczos algorithm. We then discuss the approximation of
these two-dimensional lattices by narrow ladder models that can be investigated
with well-established methods for one-dimensional correlated quantum systems,
such as the density-matrix renormalization group or bosonization. 
The validity of this approach is studied first for noninteracting electrons
and then for 
a correlated wire with a Hubbard electron-electron repulsion using
quantum Monte Carlo simulations. 
While narrow ladders
cannot be used to represent wires on metallic substrates, they capture the
physics of wires on insulating substrates if at least three legs are used.
In the second part [arXiv:1704.07359], we use this approach for a detailed numerical
investigation of a wire with a Hubbard-type interaction on an insulating
substrate.
\end{abstract}

%\pacs{71.10.Fd, 71.10.Pm, 71.27.+a, 73.21.Hb}
\maketitle

\section{\label{sec:intro}Introduction}

The fascinating properties of one-dimensional (1D) electron systems have been studied 
theoretically for more than 60 years~\cite{sol79,kiess92,gruener00,baeriswyl04,essler05,giamarchi07}.
Experimentally, quasi-1D electron systems can now be realized in linear atomic wires deposited on a semiconducting
substrate~\cite{springborg07,onc08,sni10}.
For instance, a Peierls metal-insulator transition occurs in indium chains on a
Si(111) surface~\cite{sni10,jeck16},
Luttinger liquid behavior is found in gold chains on Ge(100)~\cite{blu11} as well as in
Bi chains on InSb(100)~\cite{ohts15}, 
and it is believed that linear chains of spin-polarized and localized electrons
are formed at step edges of Si(hhk) surfaces~\cite{aul13}.
Yet, the interpretation of experimental results often remains controversial.
A fundamental issue is our poor theoretical understanding of the
effects of the coupling between an atomic wire and its three-dimensional (3D)
substrate on hallmark features of 1D systems such as the Peierls instability
or Luttinger liquid behavior. 

The theory of 1D electronic systems is mostly based
on effective models for the low-energy degrees of freedom.
The goal of this approach is to understand some generic physical phenomena
within a simplified model
rather than to achieve a full description of a specific material.  
Various quantum lattice models have become {\it de facto} standards for
describing correlated electrons in quantum wires.
One example is the 1D Hubbard model~\cite{essler05}, which can describe 
several aspects of these systems such as Luttinger liquid physics,
Mott-insulating behavior, and antiferromagnetic spin-density-wave correlations.
Obviously, we must generalize these models to include the effects of the
wire-substrate coupling.

As investigations of interacting electrons on 3D lattices with complex
geometries are extremely difficult, the modeling of wire-substrate
systems by much simpler effective models appears to be a very promising route.
Asymmetric two-leg ladder systems provide a minimal model for such
systems~\cite{springborg07,das01,das02,abd15}.
One leg represents the atomic wire, while the second leg mimics
those degrees of freedom of the substrate that couple to the wire.
For instance, this approach was used to study the stability of Luttinger liquids~\cite{das01,das02}
and Peierls insulators~\cite{springborg07} coupled to an environment.
The main advantage of these ladder models is that one can study them with
well-established methods for 1D correlated systems such as
the numerical density-matrix renormalization group (DMRG)~\cite{whi92,whi93,sch05,jec08a}
or field-theoretical techniques (e.g., bosonization and the renormalization group)~\cite{sol79,giamarchi07,Schoenhammer,Gogolin,Tsvelik}.
A significant drawback is that a single leg may be insufficient to represent 
the role of the 3D substrate~\cite{abd15}.
Until now, this approach has not been pursued systematically.

\begin{figure}[tb]
\includegraphics[width=0.24\textwidth]{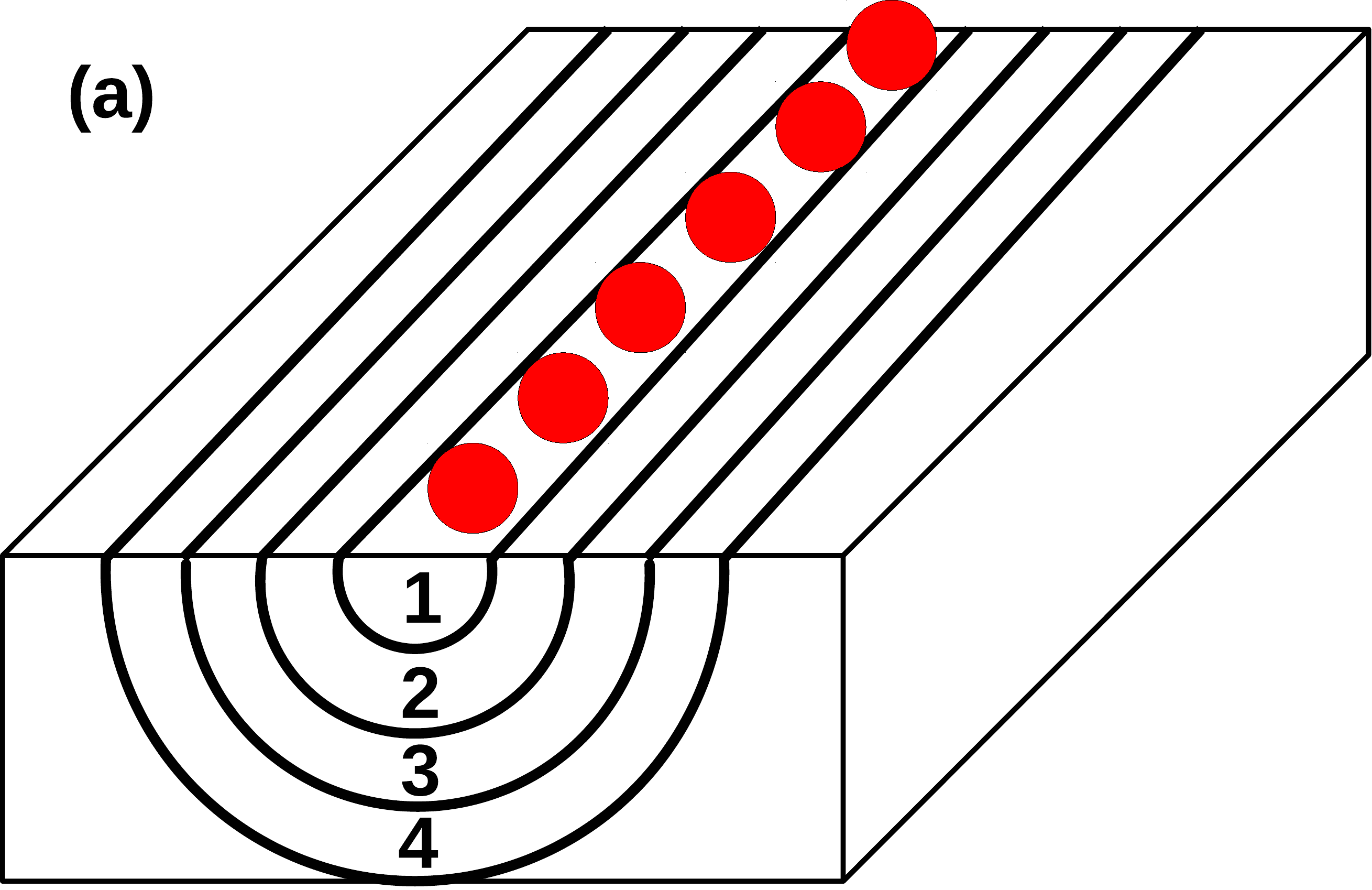}
\includegraphics[width=0.16\textwidth]{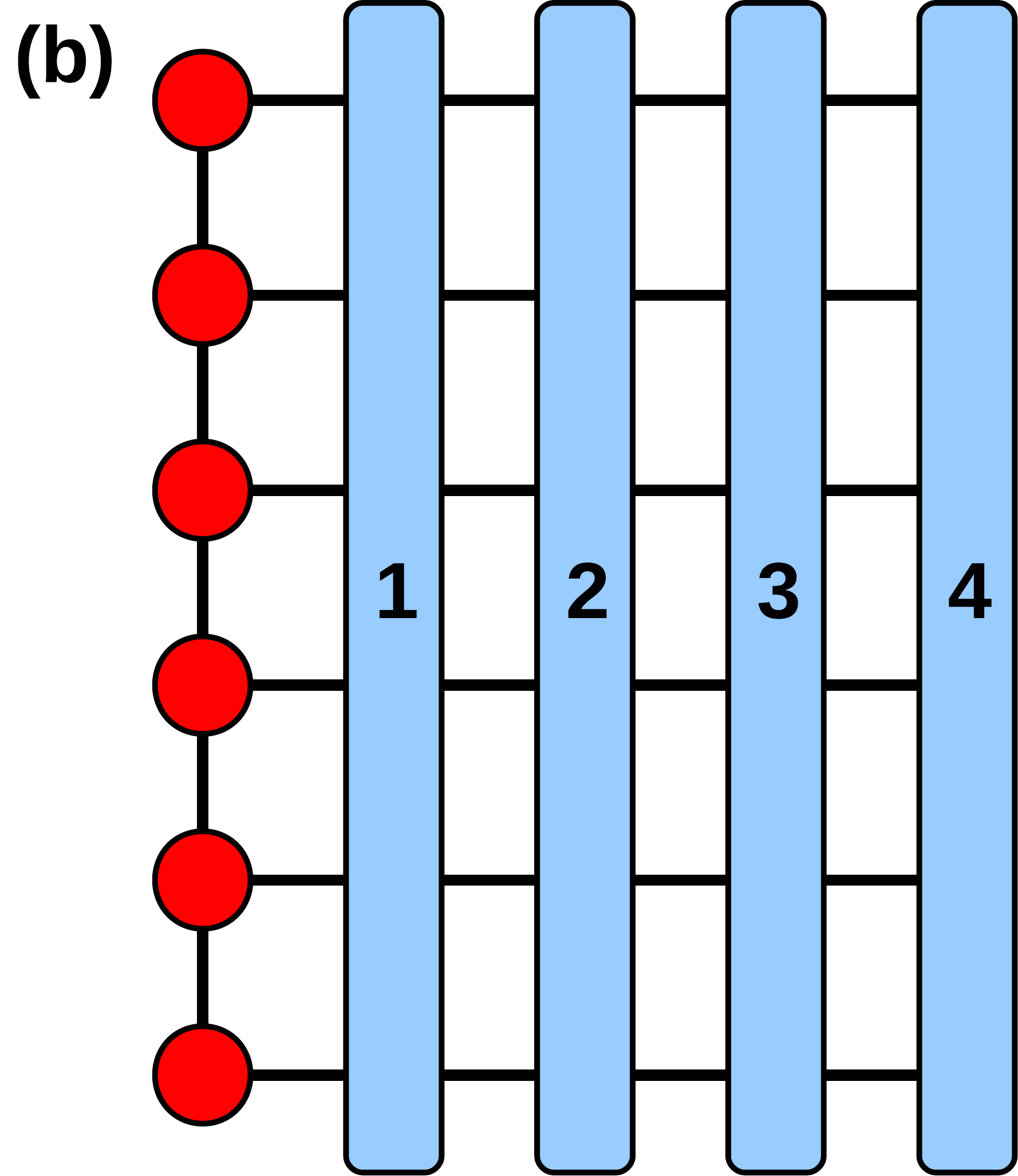}
\caption{\label{fig:mapping} (Color online) 
(a) Sketch of an atomic wire (red spheres) on a 3D substrate with four numbered shells.
(b) Ladder representation of the same system with the left-most leg corresponding
to the atomic wire (red circles) and the other legs (in blue) representing shells one to four.
}
\end{figure}

In this paper, we show how to systematically construct effective quasi-1D ladder models for wire-substrate systems.
Our method generalizes an approach recently proposed to map multi-orbital, multi-site  quantum impurity problems onto ladder systems~\cite{shi14,alle15}.
We show that if the wire-substrate system is translationally invariant in the
wire direction, it can be mapped exactly onto a wide ladder-like lattice
[i.e., an anisotropic (semi-infinite) two-dimensional (2D) lattice]. 
This mapping is illustrated in Fig.~\ref{fig:mapping}.
The key idea is to decompose the full system into independent single-impurity systems using the momentum-space
representation in the wire direction, then to perform the usual transformation of each impurity system
into a long chain~\cite{wil75,mattis81}, which finally becomes one rung of a wide ladder after transforming
back to real space in the wire direction.
The main difference between Refs.~\cite{shi14,alle15} and our method is that in the former 
the number of impurity sites and orbitals determines the ladder width and the number of single-particle host states
sets the ladder length. In contrast, in our approach, the wire length
determines the ladder length and the number of single-particle substrate
states sets the ladder width.

If the number of ladder legs (i.e., the number of shells in the substrate) is
small, we obtain a quasi-1D problem that can be treated efficiently by
methods for 1D correlated electron systems.
Thus, we investigate the approximation of the full wire-substrate system in its ladder representation
by narrow effective ladders. 
We find the approximation to be valid for an insulating substrate if at least three legs are kept,
but never for a metallic substrate.
To illustrate our procedure, we introduce an electronic 3D lattice model describing an interacting
wire on a noninteracting substrate. The procedure is demonstrated
explicitly first for a noninteracting wire and  then for 
a correlated Hubbard wire using quantum Monte Carlo (QMC) simulations~\cite{rubt05,gull11}.
In \cite{paper2}, we use DMRG and QMC
methods to investigate in detail the case of a wire with a Hubbard-type
interaction on an insulating substrate.

This paper is structured as follows. In Sec.~\ref{sec:model}, we introduce
3D lattice models for wire-substrate systems. The exact mapping of these
models onto wide ladders is presented in Sec.~\ref{sec:mapping}. In
Sec.~\ref{sec:ladder}, we discuss their approximation by effective narrow 
ladders. Finally, Sec.~\ref{sec:Conclusion} contains our conclusions.

\section{\label{sec:model}Wire-substrate model}

In this section, we introduce a model for a single correlated wire on the surface of a noninteracting 3D substrate,
which we will use in Sec.~\ref{sec:mapping} to illustrate the mapping to ladder systems.
Interactions with other wires are assumed to be negligible for the low-energy
physics. While this is not justified  for phases with long-range order,
such as Peierls, spin-density-wave, and charge-density-wave states, 
it is a reasonable approximation for Luttinger liquid phases or paramagnetic
Mott insulators, which are our main concern here. 
The system Hamiltonian can be decomposed into three terms,
\begin{equation}
\label{eq:hamiltonian}
H = H_{\text s} + H_{\text w} + H_{\text{ws}}\,,
\end{equation}
where $H_{\text s}$ describes the substrate degrees of freedom, $H_{\text w}$ the wire degrees of freedom,
and $H_{\text{ws}}$ the coupling between wire and substrate.
Although the present approach can be
applied to models that also include lattice degrees of freedom (phonons),
we focus on purely electronic models and discuss possible extensions at the
end of this section. We set $\hbar=1$ and do not distinguish between momentum
and wave number. 

\subsection{Substrate \label{sec:substrate}}

The substrate is represented by a cubic lattice with lattice constants $a=b=c=1$.
The coordinate axes are set by the lattice primitive vectors. Thus the lattice
sites have positions $\bm{r} = (x,y,z)$ with $x,y,z \in {\mathbb Z}$.
The substrate extends over $L_x, L_y$ and $L_z$ sites in the $x,y$ and $z$-directions, respectively.
We use periodic boundary conditions in the $x$ and $y$-directions but open boundary conditions 
in the $z$-direction.
The substrate surface lies in the $xy$-plane and the surface layer corresponds to $z=1$.
Thus, objects on the surface have a coordinate $z=0$.

We first introduce the simpler Hamiltonian for a metallic substrate and then generalize
it to case of an insulating substrate.
(There is at least a theoretical interest in 1D atomic structures on metallic substrates~\cite{igna10}.)
A simple model for the electronic degrees of freedom of a metallic substrate is given by a tight-binding Hamiltonian
with onsite potential $\epsilon_{\text s}$ and nearest-neighbor hopping $t_{\text s}$,
\begin{equation}
\label{eq:real-metal-sub}
H_{\text s}   =  \epsilon_{\text s} \sum_{\bm{r}, \sigma} n_{{\text s} \bm{r}  \sigma} 
- t_{\text s} \sum_{\langle \bm{r} \bm{q} \rangle} \sum_{\sigma}  \left (
c^{\dag}_{{\text s} \bm{r} \sigma}  c^{\phantom{\dag}}_{{\text s}\bm{q}\sigma} + \text{H.c.}
\right )\,.
\end{equation}
The first sum runs over all lattice sites and the second one over all
pairs $\langle \bm{r} \bm{q} \rangle$ of nearest-neighbor sites. The operator $c^{\dag}_{{\text s} \bm{r} \sigma}$ creates an electron
with spin $\sigma$ on the site with coordinates $\bm{r} = (x,y,z)$,
and the density operator for electrons with spin $\sigma$ is
$n_{{\text s} \bm{r} \sigma} = c^{\dag}_{{\text s} \bm{r} \sigma}
c^{\phantom{\dag}}_{{\text s} \bm{r}
  \sigma}$. Hamiltonian~(\ref{eq:real-metal-sub}) can be diagonalized by
the usual canonical transformation to momentum space,
\begin{equation}
\label{eq:canonical1}
d^{\dag}_{{\text s} \bm{k} \sigma}  = \sum_{\bm{r}}
\psi_{\bm{k}}(\bm{r}) \ c^{\dag}_{{\text s} \bm{r} \sigma}\,,
\end{equation}
with the single-particle eigenstates
\begin{equation}
\label{eq:wavefunction}
\psi_{\bm{k}}(\bm{r}) = \frac{1}{\sqrt{L_x}}  \ e^{i k_x x} \frac{1}{\sqrt{L_y}} \  e^{i k_y y}  \sqrt{\frac{2}{L_z+1}}
\sin(k_z z)\,.
\end{equation}
The inverse transformation is then given by
\begin{equation}
\label{eq:canonical2}
c^{\dag}_{{\text s} \bm{r} \sigma}  = \sum_{\bm{k}}
\psi^*_{\bm{k}}(\bm{r}) \ d^{\dag}_{{\text s} \bm{k} \sigma}\,,
\end{equation}
where the sum is over all sites $\bm{k}$ of the reciprocal lattice.
Here, $\bm{k} = (k_x,k_y,k_z)$ with
\begin{subequations}
\label{eq:reciprocal}
\begin{align}
k_x &= \frac{2 \pi}{L_x} n_x,  &&n_x \in {\mathbb Z}, \quad -\frac{L_x}{2} < n_x \leq \frac{L_x}{2}\,, \\
k_y &= \frac{2 \pi}{L_y} n_y,  &&n_y \in {\mathbb Z}, \quad -\frac{L_y}{2} < n_y \leq \frac{L_y}{2}\,,  \\
k_z &= \frac{\pi}{L_z+1} n_z,  &&n_z  \in {\mathbb Z}, \quad\quad\,\,\,\, 1 \leq n_z \leq L_z \,. 
\end{align}
\end{subequations}
The differences between the $z$-component and the other two components 
in Eqs.~(\ref{eq:wavefunction}) and~(\ref{eq:reciprocal}) reflect the
different boundary conditions. In the momentum-space representation 
Hamiltonian~(\ref{eq:real-metal-sub})  becomes diagonal,
\begin{equation}
\label{eq:momentum-metal-sub}
H_{\text s} =  \sum_{\bm{k},\sigma} \epsilon_{\text s}(\bm{k})
d^{\dag}_{{\text s} \bm{k}\sigma}
 d^{\phantom{\dag}}_{{\text s} \bm{k}\sigma}\,,
\end{equation}
with a single-electron dispersion 
\begin{equation}
\label{eq:metal-disp}
\epsilon_{\text s}(\bm{k})  = \epsilon_{\text s} - 2t_{\text s} [ \cos(k_x) + \cos(k_y) + \cos(k_z) ] .
\end{equation}

In reality, most substrates for atomic wires are not metallic but insulating or semiconducting~\cite{springborg07,onc08,sni10}.
A simple model for an insulating substrate consists of a valence band and a conduction band separated by a gap.
It can be constructed using the same lattice as above but with two orbitals
per site with different onsite energies 
$\epsilon_{\text v}$  and $\epsilon_{\text c}$.
The Hamiltonian then takes the form 
\begin{equation}
H_{\text s} = H_{\text v} + H_{\text c} \,,
\end{equation}
where the valence-band Hamiltonian 
\begin{equation}
\label{eq:real-valence-sub}
H_{\text v}   =  \epsilon_{\text v} \sum_{\bm{r}, \sigma}  n_{{\text v}\bm{r}\sigma}
-t_{\text v} \sum_{\langle \bm{r} \bm{q} \rangle} \sum_{\sigma} \left (
c^{\dag}_{{\text v}\bm{r} \sigma}  c^{\phantom{\dag}}_{{\text v}\bm{q}\sigma} + \text{H.c.} 
\right )
\end{equation}
and the conduction-band Hamiltonian
\begin{equation}
\label{eq:real-conduction-sub}
H_{\text c} = \epsilon_{\text c} \sum_{\bm{r}, \sigma} n_{{\text c}\bm{r}\sigma}
-t_{\text c} \sum_{\langle \bm{r} \bm{q} \rangle} \sum_{\sigma}  \left (
c^{\dag}_{{\text c}\bm{r} \sigma}  c^{\phantom{\dag}}_{{\text c}\bm{q}\sigma} + \text{H.c.}
\right )
\end{equation}
are  tight-binding Hamiltonians with nearest-neighbor hopping terms $t_{\text v}$
and $t_{\text c}$, respectively.
Accordingly, $c^{\dag}_{{\text v}\bm{r} \sigma}$ and $c^{\dag}_{{\text c}\bm{r}
\sigma}$ create electrons
with spin $\sigma$ on site $\bm{r}$ in the localized orbitals corresponding to 
the valence and conduction bands, 
while $n_{{\text v}\bm{r}\sigma}$ and $n_{{\text c}\bm{r}\sigma}$ denote
the corresponding density operators.
The canonical transformation from real to momentum space~(\ref{eq:canonical1}) can be generalized to diagonalize these
Hamiltonians. This leads to
\begin{equation}
\label{eq:momentum-valence-sub}
H_{\text v} =  \sum_{\bm{k},\sigma} \epsilon_{\text v}(\bm{k}) \ d^{\dag}_{{\text v}\bm{k}\sigma}
 d^{\phantom{\dag}}_{{\text v}\bm{k}\sigma}
\end{equation}
and 
\begin{equation}
\label{eq:momentum-conduction-sub}
H_{\text c} =  \sum_{\bm{k},\sigma} \epsilon_{\text c}(\bm{k}) \  d^{\dag}_{{\text c}\bm{k}\sigma}
 d^{\phantom{\dag}}_{{\text c}\bm{k}\sigma}
\end{equation}
with single-electron dispersions $\epsilon_{\text v}(\bm{k})$ and
$\epsilon_{\text c}(\bm{k})$ of the form~(\ref{eq:metal-disp}) but with $\{\epsilon_{\text s},t_{\text{s}}\}$
replaced by $\{\epsilon_{\text v}, t_{\text{v}}\}$ and $\{\epsilon_{\text c},t_{\text{c}}\}$, respectively.
The (possibly indirect) gap between the bottom of the conduction band and the
top of the valence band is $\Delta_{\text s} =  \epsilon_{\text c} - \epsilon_{\text v} - 6 \left (  \vert t_{\text v} \vert + \vert t_{\text c} \vert \right )$
and the condition $\Delta_{\text s} \geq 0$ restricts the range of allowed model parameters.

\subsection{Wire \label{sec:wire}}

The atomic wire is represented by a 1D chain aligned with the
$x$-direction on the substrate surface. 
To simplify the problem as much as possible, we assume that the
wire extends over the full length of the substrate, that 
the lattice constants of wire and substrate are equal,
and that every site of the wire lies exactly above the corresponding
substrate site. Thus the $L_x$ wire sites have positions $\bm{r} = (x,y_0,0)$ 
with $x=1,\dots,L_x$ and a fixed $y_0 \in \{1,\dots,L_y\}$.

The 1D Hubbard model~\cite{essler05} describes the effects of electronic correlations on the low-energy 
properties of 1D lattice systems. 
It is integrable and has been solved using the Bethe ansatz method. 
Its ground state for repulsive interactions is a Mott insulator at
half-filling  but a paramagnetic 1D metal with the low-energy properties of a Luttinger liquid
away from half-filling \cite{giamarchi07}.  Here, we use it to model the atomic wire.
The Hamiltonian is
\begin{eqnarray}
\label{eq:real-hubbard}
H_{\text w} &=&
\epsilon_{\text w} \sum_{x,\sigma} n_{{\text w} x \sigma} 
-t_{\text w}  \sum_{x,\sigma} \left ( c^{\dag}_{{\text w} x\sigma}  
c^{\phantom{\dag}}_{{\text w},x+1,\sigma} + \text{H.c.} \right ) \nonumber \\
&& + U \sum_x n_{{\text w} x \uparrow} n_{{\text w} x \downarrow}\,,
\end{eqnarray}
where $x$ runs over all wire sites, $c^{\dag}_{{\text w} x\sigma}$
creates an electron with spin $\sigma$ on the wire site at $\bm{r} = (x,y_0,0)$,
and the density operator for electrons with spin $\sigma$ is  
$n_{{\text w} x\sigma} = c^{\dag}_{{\text w} x\sigma} c^{\phantom{\dag}}_{{\text w} x\sigma}$. 
The Hubbard term of strength $U$ describes the repulsion between
two electrons on the same site, $t_{\text w}$ is the usual hopping term between
nearest-neighbor sites, and $\epsilon_{\text w}$ is the onsite potential. 

The momentum-space representation of Eq.~(\ref{eq:real-hubbard}) is  
\begin{eqnarray}
\label{eq:momentum-hubbard}
H_{\text w} &=&  \sum_{k,\sigma} \epsilon_{\text w}(k) d^{\dag}_{{\text w}k\sigma}
 d^{\phantom{\dag}}_{{\text w}k\sigma}  \\
 &&+ \frac{U}{L_x} \sum_{k,p,k',p'} d^{\dag}_{{\text w}k\uparrow}
  d^{\phantom{\dag}}_{{\text w}p\uparrow} 
 d^{\dag}_{{\text w}k'\downarrow} 
 d^{\phantom{\dag}}_{{\text w}p'\downarrow}
 \delta^{\phantom{\dag}}_{k-p,p'-k'} \nonumber
\end{eqnarray}
with the single-electron dispersion 
\begin{equation}
\label{eq:wire-disp}
\epsilon_{\text w}(k)  = \epsilon_{\text w} - 2t_{\text w} \cos(k) .
\end{equation}
The canonical transformation between real and momentum space is given by 
\begin{equation}
\label{eq:canonical3}
d^{\dag}_{{\text w} k \sigma}  = \frac{1}{\sqrt{L_x}} \sum_{x}
e^{ikx} \ c^{\dag}_{{\text w} x \sigma} .
\end{equation}
The operator $d^{\dag}_{{\text w} k \sigma}$ creates an electron with spin
$\sigma$ in an orbital with momentum $k$ in the $x$-direction and at position $(y_0,0)$ 
in the $yz$-plane.
In the above equations the indices $k,p,k'$, and $p'$ denote momenta in the
$x$-direction, see Eq.~(\ref{eq:reciprocal}a).
It is important to understand that in the framework of the 3D wire-substrate model
the above form of the Hubbard Hamiltonian corresponds to a mixed 
real-space/momentum-space representation. 

\subsection{Wire-substrate hybridization \label{sec:coupling}}

The simplest coupling between the wire and the substrate consists of a 
hybridization of the electronic orbitals. This can be realized with a hopping
term between nearest-neighbor pairs of sites located in the wire and the
substrate, respectively.

For a metallic substrate we define
\begin{equation}
\label{eq:real-hybridization-metal}
H_{\text{ws}} =  
-t_{\text{ws}}  \sum_{x,\sigma} \left ( c^{\dag}_{{\text s} \bm{r} \sigma}  
c^{\phantom{\dag}}_{{\text w} x \sigma} + \text{H.c.}  \right )
\end{equation}
with $\bm{r} = (x,y_0,1)$.
In the momentum-space representation this yields
\begin{equation}
\label{eq:momentum-hybridization-metal}
H_{\text{ws}} = \sum_{\bm{k}\sigma}  \left [ \Gamma_{\text{ws}}(\bm{k}) \
d^{\dag}_{{\text s} \bm{k} \sigma}  
d^{\phantom{\dag}}_{{\text w} k_x \sigma} + \text{H.c.} \right ]
\end{equation}
with a $k_x$-independent hybridization function 
\begin{equation}
\label{eq:hybridization-metal}
\Gamma_{\text{ws}}(\bm{k}) = - t_{\text{ws}} \sqrt{\frac{2}{L_y (L_z+1)}} \exp(-ik_yy_0) \sin(k_z)\,.
\end{equation}
For an insulating substrate with two orbitals per site the 
hybridization strengths can be different for the valence and conduction
bands, and we define 
\begin{equation}
\label{eq:hybridization-insulator}
H_{\text{ws}} = H_{\text{wv}} + H_{\text{wc}}
\end{equation}
with the hybridization between wire and valence band
\begin{equation}
\label{eq:real-hybridization-valence}
H_{\text{wv}}   =  
-t_{\text{wv}} \sum_{x,\sigma} \left ( c^{\dag}_{{\text v} \bm{r} \sigma}  
c^{\phantom{\dag}}_{{\text w} x \sigma} + \text{H.c.}  \right )
\end{equation}
and the hybridization between wire and conduction band
\begin{equation}
\label{eq:real-hybridization-conduction}
H_{\text{wc}}   =  
-t_{\text{wc}}  \sum_{x,\sigma} \left ( c^{\dag}_{{\text c} \bm{r} \sigma}  
c^{\phantom{\dag}}_{{\text w} x \sigma} + \text{H.c.}   \right ) .
\end{equation}
In the momentum-space representation, $H_{\text{wv}}$ and $H_{\text{wc}}$ are given 
by expressions similar to Eq.~(\ref{eq:momentum-hybridization-metal}),
\begin{equation}
\label{eq:momentum-hybridization-valence}
H_{\text{wv}} = \sum_{\bm{k}\sigma}  \left [ \Gamma_{\text{wv}}(\bm{k}) \
d^{\dag}_{{\text v} \bm{k} \sigma}  
d^{\phantom{\dag}}_{{\text w} k_x \sigma} + \text{H.c.} \right ]\,,
\end{equation}
\begin{equation}
\label{eq:momentum-hybridization-conduction}
H_{\text{wc}} = \sum_{\bm{k}\sigma}  \left [ \Gamma_{\text{wc}}(\bm{k}) \
d^{\dag}_{{\text c} \bm{k} \sigma}  
d^{\phantom{\dag}}_{{\text w} k_x \sigma} + \text{H.c.} \right ]\,,
\end{equation}
with $\Gamma_{\text{wv}}(\bm{k})$ and $\Gamma_{\text{wc}}(\bm{k})$
identical to Eq.~(\ref{eq:hybridization-metal}) except for the replacement
of $t_{\text{ws}}$ by $t_{\text{wv}}$ and $t_{\text{wc}}$, respectively.

\subsection{Generalizations}

Our wire-substrate model may be extended in several ways.
For instance, the substrate properties and the wire-substrate coupling can be changed without difficulty in the momentum-space representation.
For the substrate band structure we can consider general single-particle
dispersions $\epsilon_{\text b}(\bm{k})$ ($\text{b}=\text{s},\text{v},\text{c}$) beyond the simple cosine form~(\ref{eq:metal-disp}).
For the wire-substrate coupling we can define hybridization functions $\Gamma_{\text{wb}}(\bm{k})$ ($\text{b}=\text{s},\text{v},\text{c}$)
with a more general $\bm{k}$ dependence than in Eq.~(\ref{eq:hybridization-metal}).
Other possible generalizations are the modification of the wire dispersion~(\ref{eq:wire-disp}),
or the inclusion of  inter-site electron-electron interactions in the wire Hamiltonian~(\ref{eq:real-hubbard}).

Even more complicated generalizations include multiple electronic bands for
the substrate or the wire and changes of the lattice geometry or the unit cell.
Furthermore, we can also include phonon degrees of freedom in the model, with an electron-phonon coupling in the
wire and a hybridization of wire and substrate phonon modes.

\section{Exact ladder representation\label{sec:mapping}}

In this section, we explain how the wire-substrate system can be mapped onto a wide ladder
or anisotropic 2D lattice. First, the system is cut into slices perpendicular to the wire direction
in momentum space to obtain independent impurity problems with 2D hosts.
In a second step, the impurity problems are mapped onto 1D chains.
Finally, the full system is transformed back to a ladder lattice in real space.

\subsection{Impurity subsystems\label{sec:impurity}}

We first analyze Hamiltonian~(\ref{eq:hamiltonian}) for a noninteracting wire ($U=0$).
In momentum space, it can be written as a sum of independent terms,
\begin{equation}
\label{eq:hamiltonian2}
 H=\sum_{k_x,\sigma} H_{k_x\sigma}\,,
\end{equation}
with 
\begin{eqnarray}
 H_{k_x\sigma}&=&  \epsilon_{\text w}(k_x) \, d^{\dag}_{{\text w}k_x\sigma}
 d^{\phantom{\dag}}_{{\text w}k_x\sigma} 
 + \sum_{k_y,k_z} \epsilon_{\text s}(\bm{k}) \,
d^{\dag}_{{\text s} \bm{k}\sigma}
 d^{\phantom{\dag}}_{{\text s} \bm{k}\sigma} \nonumber \\
  &&+\sum_{k_y,k_z} \left[ \Gamma_{\text{ws}}(\bm{k}) \
d^{\dag}_{{\text s} \bm{k} \sigma}  
d^{\phantom{\dag}}_{{\text w} k_x \sigma} + \text{H.c.} \right]  
\label{eq:impurity-metal}
\end{eqnarray}
for a metallic substrate, or
\begin{eqnarray}
  H_{k_x\sigma}&=& \epsilon_{\text w}(k_x) \, d^{\dag}_{{\text w}k_x\sigma}
 d^{\phantom{\dag}}_{{\text w}k_x\sigma} \nonumber \\
  && +  \sum_{k_y,k_z} \epsilon_{\text c}(\bm{k}) \,
d^{\dag}_{{\text c} \bm{k}\sigma}
 d^{\phantom{\dag}}_{{\text c} \bm{k}\sigma}
 + \sum_{k_y,k_z} \epsilon_{\text v}(\bm{k}) \,
d^{\dag}_{{\text v} \bm{k}\sigma}
 d^{\phantom{\dag}}_{{\text v} \bm{k}\sigma} \nonumber \\
  && +\sum_{k_y,k_z} \left[ \Gamma_{\text{wc}}(\bm{k}) \
d^{\dag}_{{\text c} \bm{k} \sigma}  
d^{\phantom{\dag}}_{{\text w} k_x \sigma} + \text{H.c.} \right] \nonumber \\
&&+ \sum_{k_y,k_z} \left[ \Gamma_{\text{wv}}(\bm{k}) \
d^{\dag}_{{\text v} \bm{k} \sigma}  
d^{\phantom{\dag}}_{{\text w} k_x \sigma} + \text{H.c.} \right]
\label{eq:impurity-insulator}
\end{eqnarray}
for an insulating substrate. 

In either case we have $\left[H_{k_x\sigma},H_{k_x^{\prime}\sigma^{\prime}}\right]=0$\,
$\forall k_x,k_x^{\prime},\sigma,\sigma^{\prime}$. Therefore, each Hamiltonian $H_{k_x\sigma}$ can be diagonalized
and discussed separately. In the absence of electron-electron interactions,
it corresponds to a single-particle Hamiltonian acting on $N_{\text{imp}}$ sites, with
$N_{\text{imp}}=L_yL_z+1$ for a metallic substrate and
$N_{\text{imp}}=2L_yL_z+1$ for an insulating substrate.
Each $H_{k_x\sigma}$ describes a nonmagnetic impurity (the wire site with momentum $k_x$)
coupled to a two-dimensional homogeneous host [the $(k_y,k_z)$-slice of the substrate for a given $k_x$]. 
The impurity energy level is $\epsilon_{\text w}(k_x)$. For a given $k_x$
the host energies lie 
in a band between the minimum and maximum of $\epsilon_{\text s}
(\bm{k})$ for a metallic substrate, and in a band between the 
extrema of $\epsilon_{\text{c}} (\bm{k})$ and $\epsilon_{\text{v}} (\bm{k})$ for an insulating substrate.
For each $k_x$ the coupling between impurity and host is described by
the hybridization functions $\Gamma_{\text{wb}}(\bm{k})$. 
 
\begin{figure}
\includegraphics[width=0.4\textwidth]{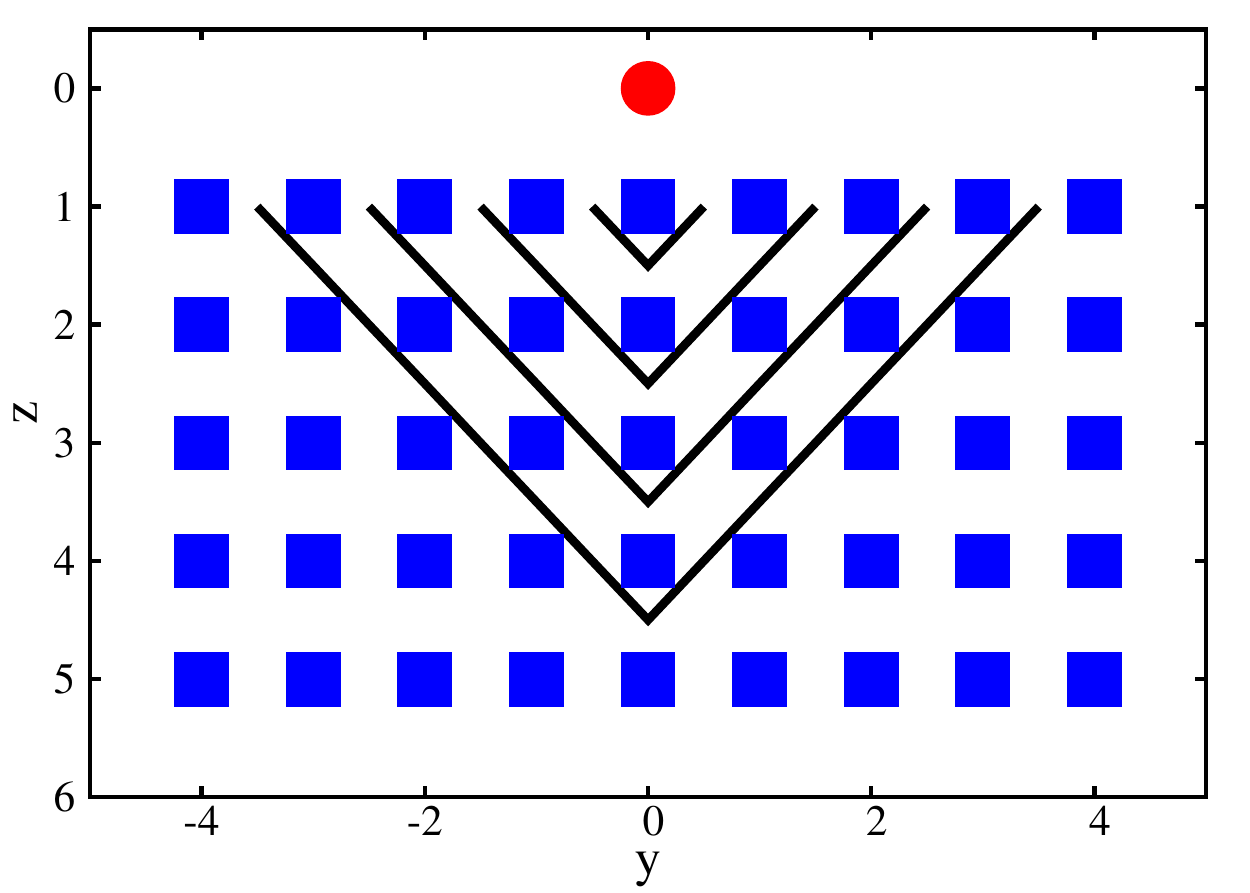}
\caption{\label{fig:impurity} (Color online)
Sketch of the impurity subsystem in the mixed representation
with $L_y=9$ and $L_z=5$.
The wire site (impurity) is represented by the red circle at $(y_0=0,z=0)$ while the substrate sites (host)
correspond to the blue squares. Black lines indicate the first, second,
third, and fourth shells (from top to bottom) around the impurity.
}
\end{figure}

The Hamiltonian $H$ can also be written in a mixed representation
combining momentum space in one direction
($k_x$) and real space in the other two directions ($y,z$).
In this representation, each $H_{k_x\sigma}$  acts on a $(y,z)$-slice of the wire-substrate system for the given 
wave vector in the $x$-direction, as illustrated in Fig.~\ref{fig:impurity}.

If the substrate slice (host) is infinitely large ($N_{\text{imp}}\rightarrow \infty$), the single-particle eigenenergies 
of each $H_{k_x\sigma}$ form one (metallic substrate) or two (insulating substrate) continua.
It is well known~\cite{Mahan} that an eigenenergy either lies in a continuum and the corresponding eigenstate is delocalized in the
impurity-host system, or the eigenenergy lies outside any continuum and the eigenstate is localized around the impurity.
In the 3D wire-substrate system the former case corresponds to states delocalized in the substrate while the latter 
corresponds to states localized in or around the wire.

\subsection{Chain representation \label{sec:chain}}

Each impurity Hamiltonian $H_{k_x\sigma}$ can be mapped onto a 1D tight-binding chain with diagonal terms and nearest-neighbor hoppings
only using the usual procedure~\cite{wil75,mattis81} based on the Lanczos tridiagonalization algorithm.
The procedure is initialized with the single-electron states $\vert k_x,-1,\sigma\rangle=0$ and 
$\vert k_x,0,\sigma\rangle=d_{\text{w}k_x\sigma}^{\dag} \vert \emptyset \rangle$, where $\vert \emptyset \rangle$ is
the vacuum state. The orthogonal one-particle states $\vert k_x,n,\sigma\rangle$ for $n=1,
\dots, N_{\text{imp}}-1$ are then generated iteratively 
using 
\begin{eqnarray}
 \vert k_x, n+1,\sigma\rangle & = & H_{k_x\sigma} |k_x,n,\sigma\rangle - A_{n}(k_x) \vert k_x,n,\sigma \rangle \nonumber \\
 && -  B^{2}_{n}(k_x) \vert k_x, n-1, \sigma \rangle\,,
\end{eqnarray}
where the coefficients are given by 
\begin{equation}
 A_{n}(k_x)=\frac{\langle k_x,n,\sigma  \vert H_{k_x\sigma} \vert k_x,n,\sigma \rangle}{\langle k_x,n,\sigma  \vert k_x,n,\sigma\rangle}
\end{equation}
for $n=0, \dots, N_{\text{imp}}-1$ and by
\begin{equation}
 B^{2}_{n}(k_x)=\frac{\langle k_x,n,\sigma \vert k_x, n,\sigma  \rangle}{\langle k_x,n-1,\sigma \vert k_x, n-1,\sigma \rangle}
\end{equation}
for $n=1, \dots, N_{\text{imp}}-1$ with $B_{0}(k_x)=0$. In the Lanczos (or
chain) representation, the Hamiltonian takes the form
\begin{eqnarray}
\label{eq:lanczosrep}
H_{k_x\sigma}&=& 
\sum^{N_{\text{imp}}-1}_{n=0}
A_{n}(k_x)f^{\dag}_{k_x n\sigma}f^{\phantom{\dag}}_{k_x n\sigma} \\
&&+\sum^{N_{\text{imp}}-2}_{n=0} \left [
B_{n+1}(k_x) f^{\dag}_{k_xn\sigma\phantom{(}} f^{\phantom{\dag}}_{k_x,n+1,\sigma}
+\text{H.c.} \right]\,,  \nonumber
\end{eqnarray}
where the new fermion operators $f^{\dag}_{k_xn\sigma}$ 
create electrons in the states $\vert k_x,n,\sigma \rangle$.
Since the wire-substrate model is invariant under spin rotation, the transformation from
the old to the new fermion operators as well as the coefficients $A_n$ and $B_n$ do not depend on spin.

This canonical transformation can be carried out numerically even when $N_{\text{imp}}$ is as large as $10^4$.
We note that the wire states are not modified by the transformation, i.e., $f^{\dag}_{k_x,n=0,\sigma} =d_{\text{w}k_x\sigma}^{\dag}$.
Moreover, one can easily show that $A_0(k_x)=\epsilon_{\text w}(k_x)$ and 
\begin{equation}
B^2_1(k_x) =  
\sum_{k_y,k_z}
 \lvert \Gamma_{\text {ws}}\left(\bm{k}\right)\rvert^{2}
\end{equation}
for the metallic substrate, while 
\begin{equation}
B^2_1(k_x)= 
\sum_{k_y,k_z}\left[
 \lvert \Gamma_{\text {wc}}\left(\bm{k}\right)\rvert^{2}
 +
 \lvert \Gamma_{\text {wv}}\left(\bm{k}\right)\rvert^{2}\right]
 \end{equation}
for the insulating substrate.

If we take the dispersion~(\ref{eq:metal-disp}) and the hybridization~(\ref{eq:hybridization-metal})
for the metallic substrate, the impurity system is relatively simple in the mixed representation of Fig.~\ref{fig:impurity}.
The impurity onsite potential is $\epsilon_{\text w}(k_x)$ while it is $\epsilon_{\text s} -t_{\text s} \cos(k_x)$ for the host sites.
A hopping $t_{\text ws}$ between the impurity and the nearest host site is the only coupling between impurity and host.
Moreover, the host sites are coupled by nearest-neighbor hopping terms $t_{\text s}$.
The state $\vert k_x,n,\sigma \rangle$ is entirely localized in a shell including the host sites
that are $m$-th nearest neighbors of the impurity with $m\leq n$. These shells are shown in Fig.~\ref{fig:impurity}.
In addition, we can show that  $A_{n}(k_x)=\epsilon_{\text s} -2t_{\text s}
\cos(k_x)$ for $n\geq 1$. The first few off-diagonal coefficients can be
computed analytically:
\begin{subequations}
\label{eq:bn}
\begin{eqnarray}
B^2_1(k_x)&=&t^2_{\text {ws}} ,\\
B^2_{2}(k_x)&=&3t^2_{\text s}, \\
B^2_{3}(k_x)&=&\frac{11}{3}t^2_{\text s}, \\
B^2_{4}(k_x)&=&\frac{125}{33}t^2_{\text s}.
\end{eqnarray}
\end{subequations}
Similarly,  we obtain for the insulating substrate
\begin{equation*}
A_{1}(k_x)  =  \frac{t^{2}_{\text {wc}} \left [\epsilon_{\text c} - 2 t_{\text c } \cos(k_x) \right ]
 +t^{2}_{\text {wv}} \left [ \epsilon_{\text v}-2t_{\text v} \cos(k_x) \right ]}{t^{2}_{\text {wc}}+t^{2}_{\text {wv}}} 
\end{equation*}
and
\begin{equation*}
B^2_1(k_x)  =  \Gamma^2(k_x) =  t^{2}_{\text{wc}}+t^{2}_{\text{wv}}. 
\end{equation*}
If we assume that the valence and conduction bands are similar, i.e., $t_{\text v} = t_{\text c} = t_{\text s}$ and
$t^{2}_{\text{wc}}=t^{2}_{\text{wv}}=t^{2}_{\text{ws}}$,
we can show that $A_{n}(k_x)=\frac{\epsilon_{\text c} + \epsilon_{\text v}}{2} -2t_{\text s} \cos(k_x)$  for $n\geq 1$
and 
\begin{subequations}
\label{eq:bn2}
\begin{eqnarray}
B^2_1(k_x)&=& 2t^2_{\text {ws}} ,\\
B^2_{2}(k_x)&=& 3t^2_{\text s}+\left ( \frac{\epsilon_{\text c} - \epsilon_{\text v}}{2} \right )^2.
\end{eqnarray}
\end{subequations}

\subsection{Real-space representation \label{sec:fullladder}}

The full Hamiltonian~(\ref{eq:hamiltonian}) can now be written using Eq.~(\ref{eq:hamiltonian2}) and the chain representations
of $H_{k_x\sigma}$,
then transformed back into the real-space representation in the $x$-direction.
As the wire states have not been modified by the mapping of the impurity subsystem to the chain representation, 
the wire Hamiltonian $H_{\text w}$ remains unchanged. The hybridization Hamiltonian becomes
\begin{equation}
H_{\text{ws}} = \sum_{x,x^{\prime},\sigma}
\left[ \Gamma(x-x^{\prime})g^{\dag}_{x,n=1,\sigma}c^{\phantom{\dag}}_{{\text w} x' \sigma}
+\text{H.c.} \right]
\end{equation}
where we defined new fermion operators
\begin{equation}
\label{eq:shellop}
 g^{\dag}_{xn\sigma}=\frac{1}{\sqrt{L_{x}}}\sum_{k_x}e^{-ik_xx}f^{\dag}_{k_xn\sigma}
\end{equation}
that create electrons with spin $\sigma$ at position $x$ in the $n$-th shell,
and with the hopping amplitudes
\begin{equation}
 \Gamma(x)=\frac{1}{L_{x}}\sum_{q}B_1(k_x)\exp(ik_xx)
\end{equation}
between wire sites and sites in the first shell in the substrate.
The substrate Hamiltonian becomes
\begin{eqnarray}
H&=&\sum^{N_{\text {imp}}-1}_{n=1}\sum_{x x^{\prime}\sigma}
A_{n}(x-x^{\prime})g^{\dag}_{xn\sigma}g^{\phantom{\dag}}_{x^{\prime}n\sigma} \\
&&+\sum^{N_{\text{imp}}-2}_{n=1}\sum_{xx^{\prime}\sigma}
\left[ B_{n+1}(x-x^{\prime})g^{\dag}_{xn\sigma}g^{\phantom{\dag}}_{x^{\prime},n+1,\sigma}
+\text{H.c} \right] \nonumber
\end{eqnarray}
with the hopping amplitudes 
\begin{equation}
 A_{n}(x)=\frac{1}{L_{x}}\sum_{k_x}A_{n}(k_x)\exp(ik_xx)
\end{equation}
in the wire direction within the same shell $n$
(or the onsite potential for $x=0$) and the hopping amplitudes 
\begin{equation}
 B_{n+1}(x)=\frac{1}{L_{x}}\sum_{k_x}B_{n+1}(k_x)\exp(ik_xx)
\end{equation}
between sites in shells $n$ and $n+1$.
Therefore, we have obtained a new representation of the Hamiltonian $H$ with long-range hoppings on a 2D lattice of
size $L_x \times N_{\text{imp}}$.

This complex system can be simplified considerably if we assume that
the hybridization functions are independent of the $x$-component of the wave vector $\bm{k}$,
\begin{equation} 
\label{eq:condition1}
\Gamma_{\text{wb}}\left(\bm{k}\right) =  \Gamma_{\text{wb}}\left(k_y,k_z\right) \quad (\text{b=s,c,v})
\end{equation} 
and that the dispersions have the additive form 
\begin{equation}
\label{eq:condition2}
 \epsilon_{\text b}\left(\bm{k}\right)=\nu\left(k_x\right)+\epsilon_{\text b}\left(k_y,k_z\right) \quad (\text{b=s,c,v})\,.
\end{equation}
[These conditions are fulfilled for the tight-binding Hamiltonians defined in
Sec.~\ref{sec:model}, see, e.g., Eqs.~(\ref{eq:metal-disp})
and~(\ref{eq:hybridization-metal}), but $t_{\text{c}}=t_{\text v}=t_{\text s}$
is required for the insulating substrate.]
In that case, the impurity Hamiltonians $H_{k_x\sigma}$ depend on the momentum $k_x$ only through the impurity onsite potential
$\epsilon_{\text w}(k_x)$ and a constant energy shift $\nu(k_x)$ in the substrate.
Therefore, the chain representations of the substrate are identical for all wave vectors $k_x$ up to energy shifts.
It follows that the hybridization between wire and substrate is
\begin{equation}
\Gamma(x)=\Gamma\delta_{x,0}
\end{equation}
with $\Gamma=B_1(k_x)$ and that
the hopping terms between nearest-neighbor shells are
\begin{equation}
B_{n}(x)=-t^{\text{rung}}_{n}\delta_{x,0} \quad (n\geq 2)
\end{equation}
with $t^{\text{rung}}_{n}=-B_n(k_x)$.
In addition,  one finds that
\begin{equation}
A_{n}(x)=-t^{\text{leg}}_x+\mu_{n}\delta_{x,0} \quad (n\geq 1)
\end{equation}
with
\begin{equation}
t^{\text{leg}}_x=-\frac{1}{L_{x}}\sum_{k_x}\nu(k_x) \exp(ik_x x) 
\end{equation}
and $\mu_{n} = A_{n}(k_x) - \nu(k_x)$.

At this point, we have obtained a representation of the wire-substrate Hamiltonian $H$ in
the form of ladder system with $L_x$ rungs and $N_{\text{imp}}$ legs, as
sketched in Fig.~\ref{fig:mapping}. The leg with $n=0$ is the wire, in particular $g^{\dag}_{x,n=0,\sigma} = c^{\dag}_{\text{w}x\sigma}$,
while legs with $n=1,\dots,N_{\text{imp}}-1$ correspond to the successive shells and represent the substrate. 
The full Hamiltonian~(\ref{eq:hamiltonian}) consists of the unchanged wire Hamiltonian $H_{\text w}$, a hopping term $\Gamma$ (hybridization) between sites at the same position $x$
in the wire and the first leg, a nearest-neighbor rung hopping $t^{\text{rung}}_n$ between substrate legs $n-1$ and $n$,
an onsite potential $\mu_n-t^{\text{leg}}_0$ constant within each substrate leg, 
and the same intra-leg hopping terms $t^{\text{leg}}_x$ in every substrate
leg; the latter are identical to the hopping terms in the original substrate
Hamiltonian $H_{\text s}$.

Equations~(\ref{eq:condition1}) and~(\ref{eq:condition2}) are 
the main conditions on the wire-substrate system to make the
mapping possible, in addition to translation symmetry in the
$x$-direction. Although we have derived the mapping for a noninteracting wire only, it is clear that the wire sites and
their Hamiltonian $H_{\text w}$ are not modified by the transformation of the substrate.
Therefore, the mapping remains valid even if $H_{\text w}$ includes a Hubbard repulsion,
more general electron-electron interactions, or electron-phonon coupling.

\begin{figure}
\includegraphics[width=0.4\textwidth]{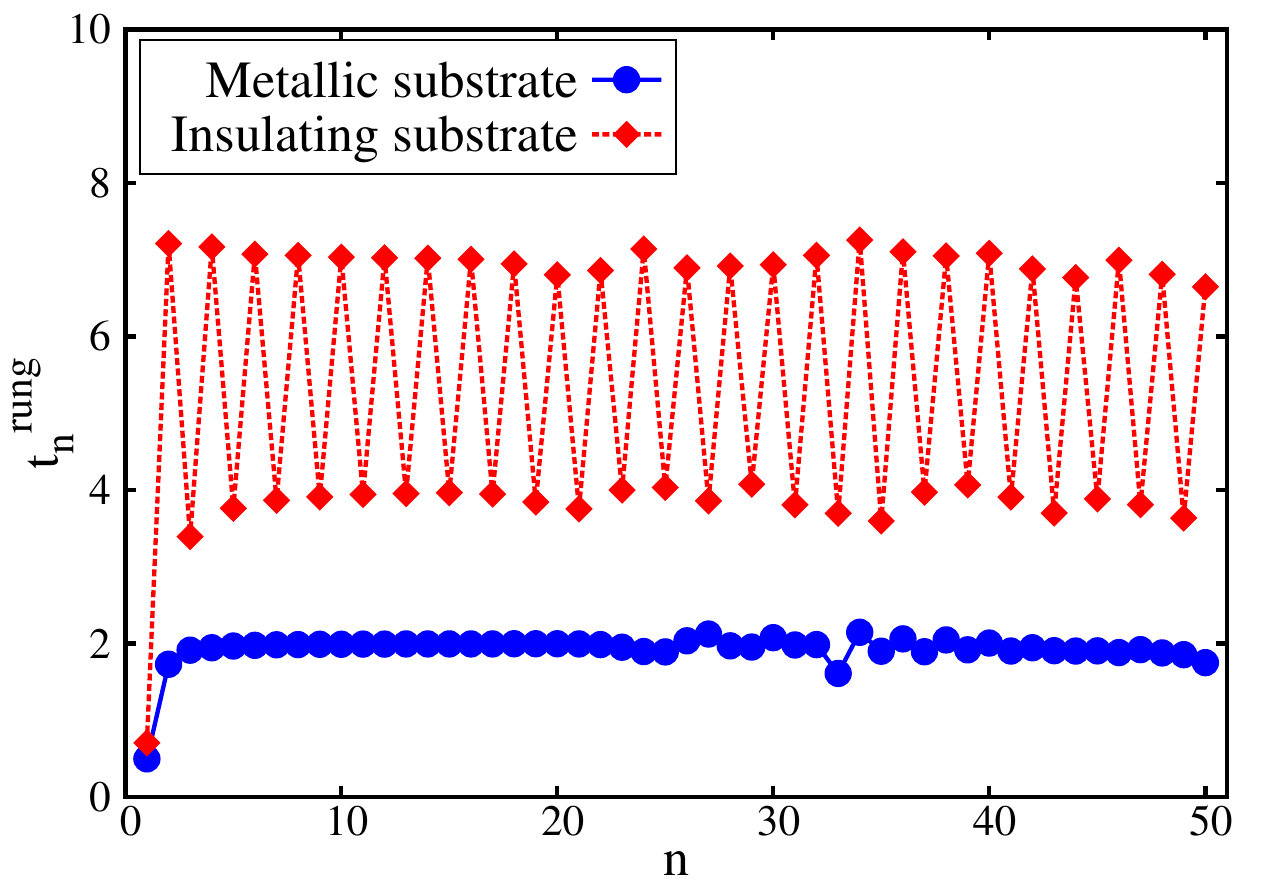}
\caption{\label{fig:hoppings} (Color online)
Hopping integrals $t^{\text{rung}}_n$ between legs $n-1$ and $n$ calculated numerically with the Lanczos algorithm
for a metallic substrate (circles, $t_s=1$, $t^{2}_{\text{ws}}=0.25$, $L_y=32$, $ L_z=16$) and an insulating substrate
(diamonds, $t_{\text{c}}=t_{\text{v}}=1$,
$t^{2}_{\text{wc}}=t^{2}_{\text{wv}}=0.25$, $\epsilon_{\text c} = -\epsilon_{\text v}= 7$, $L_y=32$, $L_z=8$).
}
\end{figure}

For substrates with dispersions of the form~(\ref{eq:metal-disp}) we have
$\nu(k_x)=-2t_{\text s} \cos(k_x)$, so that hopping within substrate legs
takes place between nearest-neighbors only,
\begin{equation}
t^{\text{leg}}_x=\begin{cases}
	        t_{\text s}  &\text{if } \lvert x \rvert=1\,, \\
		0 &\text{otherwise}\,.
                 \end{cases}
\end{equation}
The explicit form of the full Hamiltonian is then
\begin{eqnarray}
\label{eq:ladder-hamiltonian}
 H&=&H_{\text w}+\sum_{x,\sigma}\left(\Gamma\, g^{\dag}_{x,n=1,\sigma}c^{\phantom{\dag}}_{\text{w}x\sigma}+\text{H.c.}\right) \nonumber \\
  &&+\sum^{N_{\text{imp}}-1}_{n=1}\sum_{x,\sigma}\mu_{n} \, g^{\dag}_{xn\sigma}g^{\phantom{\dag}}_{xn\sigma} \\
  &&-t_{\text s} \sum^{N_{\text{imp}}-1}_{n=1}\sum_{x,\sigma} \left ( g^{\dag}_{xn\sigma}g^{\phantom{\dag}}_{x+1,n\sigma} +\text{H.c.}\right) \nonumber \\
  &&-\sum^{N_{\text{imp}}-2}_{n=1}\sum_{x,\sigma}\left( t^{\text{rung}}_{n+1} \, g^{\dag}_{xn\sigma}g^{\phantom{\dag}}_{x,n+1,\sigma}+\text{H.c.}\right). \nonumber
\end{eqnarray}
For the metallic substrate, $\mu_n=\epsilon_{\text s}$, $\Gamma=-t_{\text{ws}}$, and the first few hoppings 
$t^{\text{rung}}_{n}=-B_n(k_x)$ are given by Eq.~(\ref{eq:bn}).
For the insulating substrate with $t_{\text{c}}=t_{\text v}=t_{\text s}$ and
$t^{2}_{\text{wc}}=t^{2}_{\text{wv}}=t^{2}_{\text{ws}}$, we have
$\mu_n=(\epsilon_{\text c} + \epsilon_{\text v})/{2}$,
$\Gamma=-B_1(k_x)$, and $t^{\text{rung}}_{2}=-B_2(k_x)$ as given by Eq.~(\ref{eq:bn2}).
The hopping terms $t^{\text{rung}}_{n}$ for larger $n$ can be computed numerically with the Lanczos algorithm, as described
in Sec.~\ref{sec:chain}.
Figure~\ref{fig:hoppings} shows the hopping terms calculated for a metallic and an insulating substrate. We see that 
for large $n$
they converge to about $2t_{\text s}$  in the metallic case. The feature near $n=32$ is due to the
finite substrate size used in the calculation  ($L_y=32$ and $L_z=16$). 
In the insulating case, the hopping terms oscillate between $3t_{\text s}$ 
and about $7t_{\text s}$, as required to generate the gap $\approx \epsilon_{\text c} - \epsilon_{\text v}$
between valence and conduction bands at a fixed $k_x$ in this representation of the substrate.
Equation~(\ref{eq:bn2}) shows the different dependence of the first two hopping terms on $\epsilon_{\text c} - \epsilon_{\text v}$
explicitly.

\subsection{Alternative representations}

The condition~(\ref{eq:condition2}) is inconvenient for the insulating substrate because it imposes
the same dispersion $\nu(k_x)$ in the $k_x$-direction in the valence and conduction bands 
to obtain a ladder-like Hamiltonian. However, it is possible to overcome this
restriction by using an alternative mapping. Since the electronic states of the conduction and valence bands interact only through the impurity site, we can use
a two-chain representation of the impurity problem where one chain represents the valence-band sites
and the other the conduction-band sites. 
(A similar two-chain representation is used for the lower and upper Hubbard bands in
Mott-Hubbard insulators~\cite{florian}.) Each chain can be generated separately using
the Lanczos algorithm as described in Sec.~\ref{sec:chain}. This yields a ladder-like Hamiltonian
if condition~(\ref{eq:condition2}) is satisfied separately by the conduction and valence bands, i.e.,
the dispersions $\nu_{\text c}(k_x)$ and $\nu_{\text v}(k_x)$ can be different.
The wire is then the middle leg of the ladder and the legs representing valence and conduction bands
extend on either side. The Hamiltonian parameters $A_n(k_x)$, and $B_n(k_x)$ are given
by equations similar to those obtained for the metallic substrate, e.g., by Eq.~(\ref{eq:bn}) with
$\{\epsilon_{\text s},t_{\text s}\}$ replaced by $\{\epsilon_{\text v},t_{\text v}\}$ and
$\{\epsilon_{\text c},t_{\text c}\}$, respectively. 

The condition of a translationally invariant wire may be relaxed by performing the mapping
first for a substrate decoupled from the wire and then connecting the two
subsystems via the hybridization term. Such a generalization allows us
to consider wires with disorder, a periodic potential modulation, or a nonuniform hybridization with the substrate.
In addition, it should  be possible to apply the mapping to systems with 
an electron-electron interaction between the wire and the adjacent substrate chain.
We illustrate this procedure
for a substrate dispersion of the form~(\ref{eq:metal-disp}), written in
a mixed ($k_x,y,z$) representation. We can then perform the Lanczos iteration
starting from the site ($k_x,y_0,z=1$) to obtain a chain representation. For a
metallic substrate this gives $A_{n}(k_x)=\epsilon_{\text s} -2t_{\text s} \cos(k_x)$
for $n\geq 0$ and
\begin{subequations}
\begin{eqnarray}
B^2_{1}(k_x)&=&3t^2_{\text s}\,, \\
B^2_{2}(k_x)&=&\frac{11}{3}t^2_{\text s}\,, \\
B^2_{3}(k_x)&=&\frac{125}{33}t^2_{\text s}\,.
\end{eqnarray}
\end{subequations}
Hence, similar to the translation-invariant case, we obtain a ladder representation for the substrate by
transforming back to real-space along the $x$-direction. A general
wire-substrate hybridization can then be introduced in the form of an $x$-dependent hopping between
the wire and the adjacent leg of the effective ladder model.

\section{Effective narrow ladder models \label{sec:ladder}}

The above mapping of the wire-substrate model is exact. However, the
resulting ladder representation corresponds to an anisotropic 2D system
rather than a quasi-1D system because the number of legs is 
proportional to the number $N_{\text{imp}}$ of single-particle states in a $(yz)$-slice of the system and
thus generally very large. Nevertheless, we have at least two reasons to
believe that quasi-1D effective systems on narrow ladders can be sufficient
to accurately represent the wire-substrate system.

First, intuitively, 1D physics (such as Luttinger liquid behavior) should occur in the wire
or in a region of the substrate around the wire.
This corresponds to legs that are close to the wire in the ladder representation, see Fig.~\ref{fig:mapping}.
Thus, the legs that are distant from the wire should not be essential for a qualitative description of 1D properties.
Second, it can be shown~\cite{kniz12} that the size of an effective representation for the environment of a quantum subsystem
does not need to be larger than the size of the subsystem itself. In our problem, this implies
that, in principle,  the substrate can be represented by an effective lattice that is not larger than the wire,
i.e., by a single leg.
Unfortunately, in that case the effective Hamiltonian depends on the specific quantum state considered and 
the only known method for determining it exactly is to solve the full wire-substrate model.

Therefore, in this section, we explore the applicability of effective narrow
ladder models (NLMs) with $N_{\text {leg}}$ legs, obtained by considering only the legs closest to the
wire in the ladder representation~(\ref{eq:ladder-hamiltonian}). 
The focus will be on the single-particle densities of states and the $k_x$-resolved
single-particle spectral functions.
The  spectral function for the wire is defined in
the momentum-space representation~(\ref{eq:canonical3}) by 
\begin{eqnarray}
A_{\text{w}}(\omega,k_x) & = & 
\sum_{\alpha} 
\left \vert \left \langle \alpha \left \vert 
d^{\dag}_{{\text w} k_x \sigma} 
\right \vert 0 \right \rangle \right \vert^2 
\delta \left ( \omega - E_{\alpha}+E_0  \right ) \qquad \nonumber \\
& + & \sum_{\alpha} \left \vert \left \langle \alpha \left \vert d^{\phantom{\dag}}_{{\text w} k_x \sigma} \right \vert 0 \right \rangle \right \vert^2
\delta \left ( \omega + E_{\alpha}-E_0  \right )\,, 
\label{eq:spectral}
%\nonumber
\end{eqnarray}
where $\vert \alpha \rangle$ and $E_{\alpha}$ denote the many-body
eigenstates and eigenvalues of $H$ in its representation~(\ref{eq:ladder-hamiltonian})
with $N_{\text {leg}}$ substituted for $N_{\text {imp}}$, while
$\vert 0 \rangle$ and $E_{0}$ indicate the ground state and its energy.

In the real-space ladder representation (Sec.~\ref{sec:fullladder}),
the $k_x$-resolved spectral functions $A(\omega,k_x,n)$ are defined 
for each leg $n\geq 0$ by the same expression~(\ref{eq:spectral})
with $f^{\dag}_{k_xn\sigma}$ and $f^{\phantom{\dag}}_{k_xn\sigma}$ substituted
for $d^{\dag}_{{\text w} k_x \sigma}$ and $d^{\phantom{\dag}}_{{\text w} k_x \sigma}$.
Obviously, $A(\omega,k_x,n=0)=A_{\text{w}}(\omega,k_x)$
while $A(\omega,k_x,n\geq 1)$ relates to single-particle excitations in the substrate
and the overall spectral function for the substrate $A_{\text{s}}(\omega,k_x)$ is obtained by averaging 
over all $n\geq 1$.

Similarly, in the mixed representation $(k_x,y,z)$ (see Sec.~\ref{sec:impurity}),
the spectral function $A_{\text{s}}(\omega,k_x,y,z)$ 
is obtained
through substitution of $h^{\dag}_{{\text b}k_xyz\sigma}$ and $h^{\phantom{\dag}}_{{\text b}k_xyz\sigma}$ 
for $d^{\dag}_{{\text w} k_x \sigma}$ and $d^{\phantom{\dag}}_{{\text w} k_x \sigma}$
in the definition of $A_{\text{w}}(\omega,k_x)$, where
$b=\text{s}$ for a metallic substrate while we average over both bands  
($b=\text{v},\text{c}$) for an insulating substrate.
Here we introduced a mixed-representation fermion operator
\begin{equation}
h^{\dag}_{\text{b}k_xyz\sigma} = \frac{1}{\sqrt{L_{x}}}\sum_{x}e^{ik_xx}c^{\dag}_{{\text b}\bm{r}\sigma}
\end{equation}
that creates an electron with spin $\sigma$ and momentum $k_x$ in the wire direction
and coordinates ($y,z$) in the other directions. Hence,
$A_{\text{s}}(\omega,k_x,y,z)$ is the $k_x$-resolved spectral function 
for a substrate chain parallel to the atomic wire at position ($y,z$).
Finally, the overall spectral function for the substrate $A_{\text{s}}(\omega,k_x)$
is obtained by averaging $A_{\text{s}}(\omega,k_x,y,z)$ over all coordinates $y$ and $z$.

The densities of states (DOSs) in the wire ($\text{b}=\text{w}$) and
the substrate ($\text{b}=\text{s}$) are 
\begin{equation}
\label{eq:dos}
 D_{\text{b}}(\omega)=\frac{1}{L_x}\sum_{k_x} A_{\text{b}}(\omega,k_x)\,.
\end{equation}
With these definitions, both DOSs are normalized, i.e., their integral over
all frequencies $\omega$ equals one.
Thus, we have defined the spectral functions and DOSs of the NLM~(\ref{eq:ladder-hamiltonian})
for any $N_{\text {leg}} \leq N_{\text {imp}}$.
For $N_{\text {leg}} = N_{\text {imp}}$, we recover the spectral functions
and DOSs of the full wire-substrate system.
The normalized total DOS of the NLM is $D(\omega) = [ D_{\text{w}}(\omega) +
(N_{\text {leg}}-1) D_{\text{s}}(\omega)]/N_{\text {leg}}$, so that the
spectral weight of the wire becomes negligible compared to that of the substrate
for $N_{\text {leg}} \gg 1$. In the remainder of this section, we assess the
quality of the NLM approximation first for a noninteracting wire and then
for a correlated wire by comparing spectral functions.

\subsection{Noninteracting wire \label{sec:noninteracting}}

For noninteracting systems [i.e., $U=0$ in Eq.~(\ref{eq:real-hubbard})], we focus on comparing spectral properties
of the full system with those of the NLM with various numbers of legs $N_{\text {leg}}$.
The electron spin will be omitted in this section as it just gives an overall
factor of two. To compute spectral properties, 
we used the Hamiltonians $H_{k_x}$ in their tridiagonal Lanczos 
representations~(\ref{eq:lanczosrep}), projected onto the subspace given by
the first $N_{\text {leg}}$ Lanczos vectors, i.e., substituting $N_{\text
  {leg}}\leq N_{\text {imp}}$ for $N_{\text {imp}}$ in Eq.~(\ref{eq:lanczosrep}).
Let $\psi_{\lambda k_x}(n)$ and $\varepsilon_{\lambda k_x}$ $(\lambda=1,\dots,N_{\text {leg}})$
denote the one-particle eigenstates and eigenvalues of these Hamiltonians.  
The spectral function in this chain representation is given by 
\begin{equation}
A(\omega,k_x,n)=\sum_{\lambda=1}^{N_{\text{leg}}}
\lvert\psi_{\lambda k_x}(n)\rvert^{2}\delta\left(\omega-\varepsilon_{\lambda k_x}\right)
\end{equation}
and can be easily calculated for any $1 \leq N_{\text {leg}} \leq N_{\text {imp}}$.
As discussed above, the $k_x$-resolved spectral function for the wire is
given by
\begin{equation}
A_{\text{w}}(\omega,k_x)=A(\omega,k_x,0)\,,
\end{equation}
whereas for the substrate we have
\begin{equation}
A_{\text{s}} (\omega,k_x)=\frac{1}{N_{\text{leg}}-1} \sum_{n=1}^{N_{\text{leg}}-1}
A(\omega,k_x,n)\,.
\end{equation}

For insulating substrates, we can find model parameters such that
some single-particle eigenenergies lie in the substrate band gap.
The corresponding eigenstates are then localized on or around the wire,
i.e., the density
$\lvert\psi_{\lambda k_x}(n)\rvert^{2}$ remains finite on the wire sites ($n=0$)
or the neighboring substrate sites (small $n$) for $N_{\text {imp}} \rightarrow \infty$.
These states form a band in the $k_x$-direction within the substrate band gap.
A wave packet built from such states remains in or around the wire but can travel
freely in the wire direction. Hence, the states represent a 1D electronic
subsystem in the 3D wire-substrate system and will be our focus.
(Other cases, such as all single-particle energies inside the valence or conduction bands,
are not relevant for real wire-substrate materials.) 

\begin{figure}[t]
\hspace*{7mm}
\includegraphics[width=0.38\textwidth]{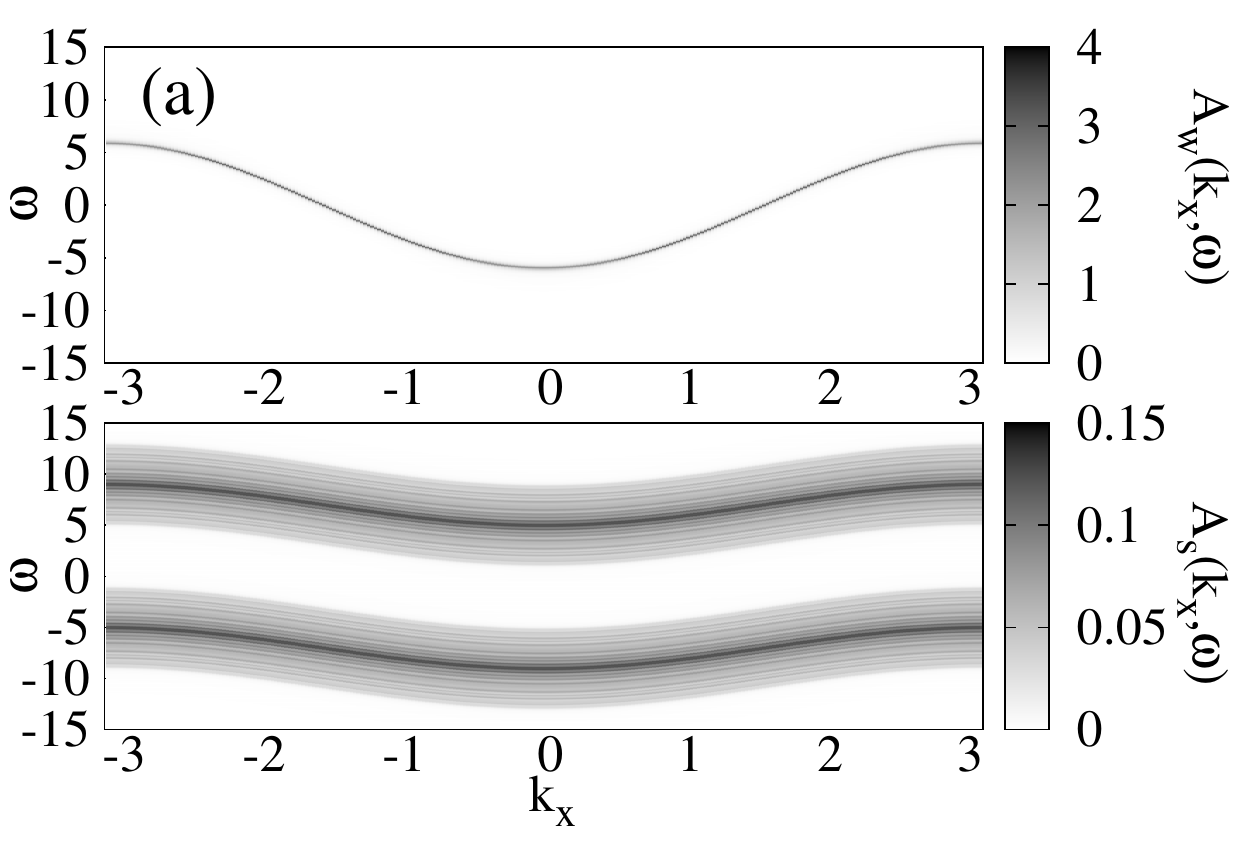}
\hspace*{-7.5mm}
\includegraphics[width=0.36\textwidth]{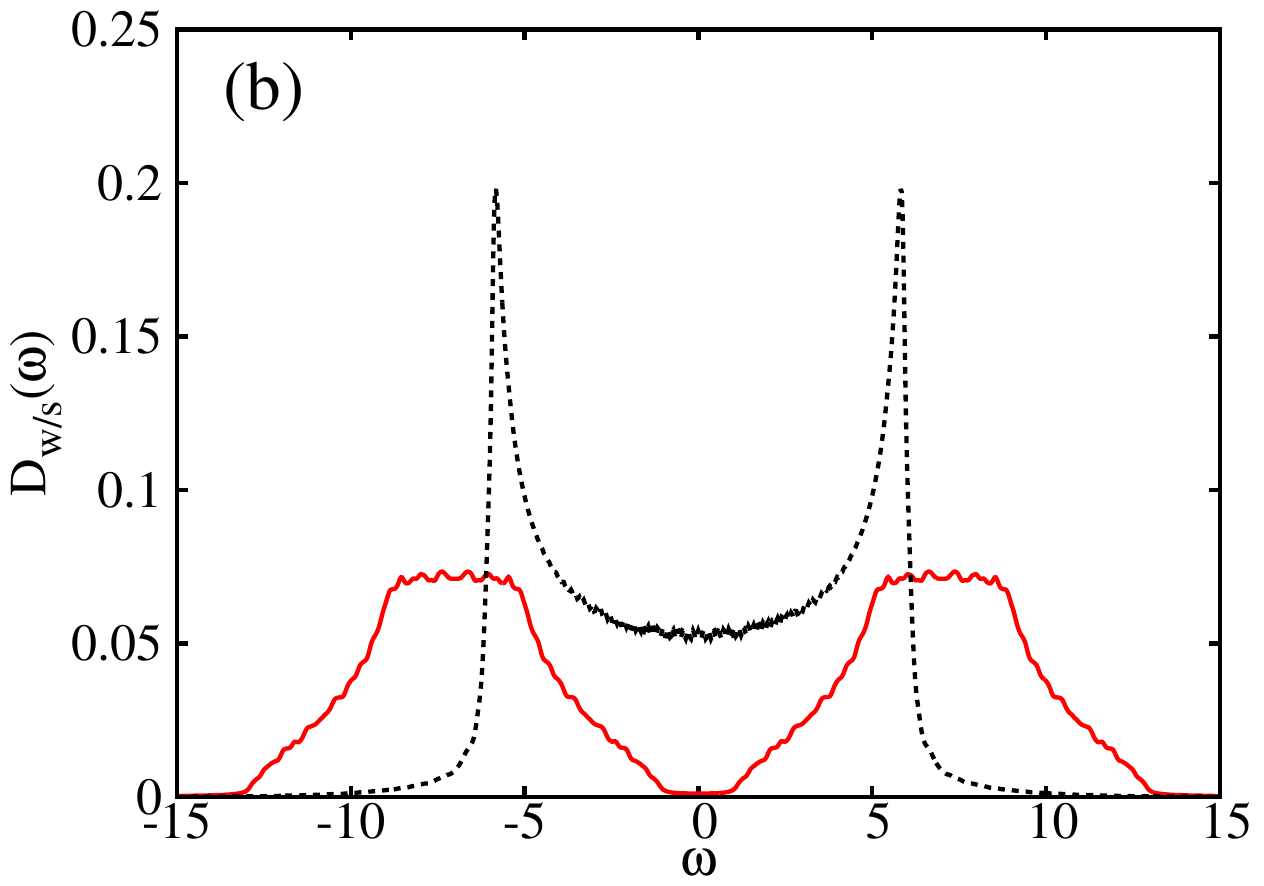}
\caption{\label{fig4} (Color online) 
Spectral properties of a noninteracting wire on an insulating 3D substrate with
$N_{\text{leg}} = N_{\text{imp}}=513$. The other parameters are given in the text.
(a) Spectral functions in the wire [$A_w(\omega,k_x)$, top] and
the substrate [$A_s(\omega,k_x)$, bottom]. (b) DOS in the wire
[$D_{\text{w}}(\omega)$, dotted black line] and in the substrate
[$D_{\text{s}}(\omega)$, solid red line].
}
\end{figure}

Figure~\ref{fig4} shows spectral functions and DOSs for a noninteracting wire on an insulating substrate.
The wire hopping is $t_{\text{w}}=3$ while the hybridization
between wire and substrate is chosen to be $t_{\text{ws}}=0.5$.
(For all examples discussed in Sec.~\ref{sec:ladder}, we will only use a symmetric wire-substrate hybridization
$t_{\text{wc}}=t_{\text{wv}}$; in the following this parameter will be denoted as $t_{\text{ws}}$.)
The substrate parameters are 
$t_{\text{c}}=t_{\text{v}}=1$ and 
$\epsilon_{\text{c}}=-\epsilon_{\text{v}}=7$.
The system sizes are $L_x=256, L_y=32$ and $L_z=8$.
Half-filling corresponds to the Fermi energy $\epsilon_{\text{F}}=0$.
These model parameters correspond to an indirect gap $\Delta_{\text{s}}=2$ 
and a constant direct gap $\Delta(k_x) = 6$ for all $k_x$
in the substrate single-particle spectrum in the absence of a wire, or for a
vanishing wire-substrate coupling
(i.e., $t_{\text{ws}}=0$).

These gaps remain visible for a nonzero wire-substrate hybridization, as illustrated
for $t_{\text{ws}}=0.5$ in Fig.~\ref{fig4}. The wire spectral weight is concentrated in a single band within the substrate band gap and crosses the Fermi energy $\epsilon_{\text{F}}\approx 0$ if the system
is at or close to half-filling. This band resembles the cosine dispersion~(\ref{eq:wire-disp}) of the uncoupled
wire but has a small intrinsic width due to the wire-substrate hybridization $t_{\text{ws}}$.
Only the wire band edges at $k_x \approx \pm \pi$ and $k_x \approx 0$  overlap with the substrate conduction and 
valence bands in the spectral functions shown in Fig.~\ref{fig4}(a).
As we can see in Fig.~\ref{fig4}(b), the DOS $D_{\text{w}}(\omega)$
of the wire retains the typical profile of a 1D tight-binding system with square-root singularities
at the band edges close to $\pm 2t_{\text w}$. In the substrate, $D_{\text{s}}(\omega)$ exhibits 
the overall shape of a 3D tight-binding system despite the fact that the DOSs
overlap over a broad energy range. This confirms that an effective quasi-1D electron system subsists in the wire 
despite the nonnegligible
wire-substrate hybridization $t_{\text{ws}}$. 
Note that the continuous but jagged DOS curves arise from finite-size effects
and a broadening of $\delta$-peaks into Lorentzians of width $\eta= 0.1$.

For a stronger hybridization $t_{\text{ws}}$, we observe three bands situated symmetrically around
the middle of the substrate band gap in the wire spectral functions.
The distance between these bands grows with $t_{\text{ws}}$ and for strong enough hybridization 
(e.g., $t_{\text{ws}}=8$) the lower and upper bands are located below the valence band and above the conduction band,
respectively.
The dispersive central band is similar to the single band found at smaller $t_{\text{ws}}$ and
shown in the upper panel of Fig.~\ref{fig4}(a).
This feature can be understood in the limit of strong hybridization 
$\epsilon_{\text c} - \epsilon_{\text v},t_{\text{ws}} \gg t_{\text{w}} \gg t_{\text{s}}$.
In first approximation, each wire site forms a trimer with the two orbitals on its first nearest-neighbor substrate site 
because of their strong effective rung hoppings, see Eq.~(\ref{eq:bn2}). 
The wire hopping $t_{\text{w}}$ then leads to the formation of a strong-rung three-leg ladder made of these same sites.
The single-particle eigenenergies of this system form three bands of width $\propto t_{\text{w}}$, which
are finally slightly hybridized with the rest of the
substrate by a weak effective coupling $\propto t_{\text{s}}$.
The eigenstates of the central band that are within the substrate band gap are localized on the wire
and its first nearest-neighbor substrate sites but delocalized in the wire direction.
Thus they form a quasi-1D electron system on and around the wire like the eigenstates
of the single band found at weaker hybridization.

\begin{figure}[t]
\hspace*{7mm}
\includegraphics[width=0.38\textwidth]{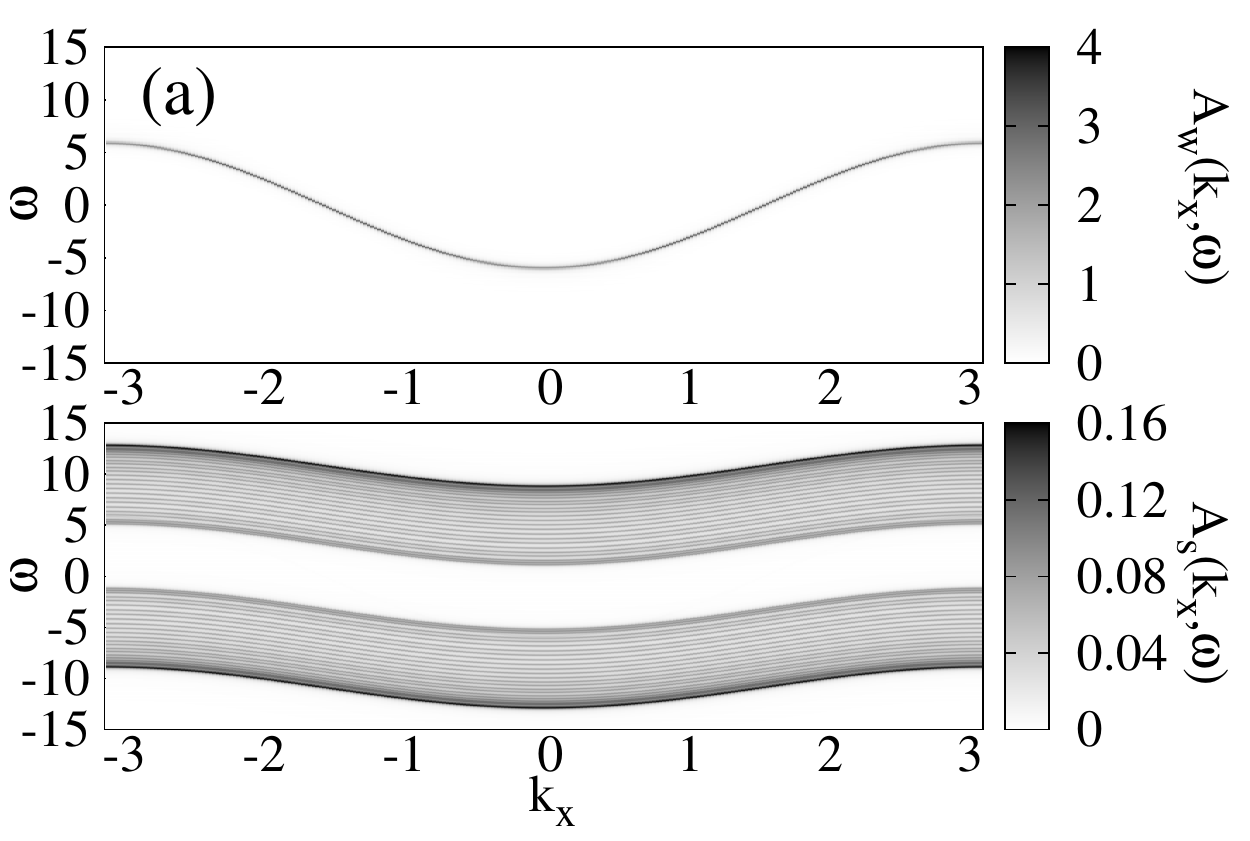}
\hspace*{-7.5mm}
\includegraphics[width=0.36\textwidth]{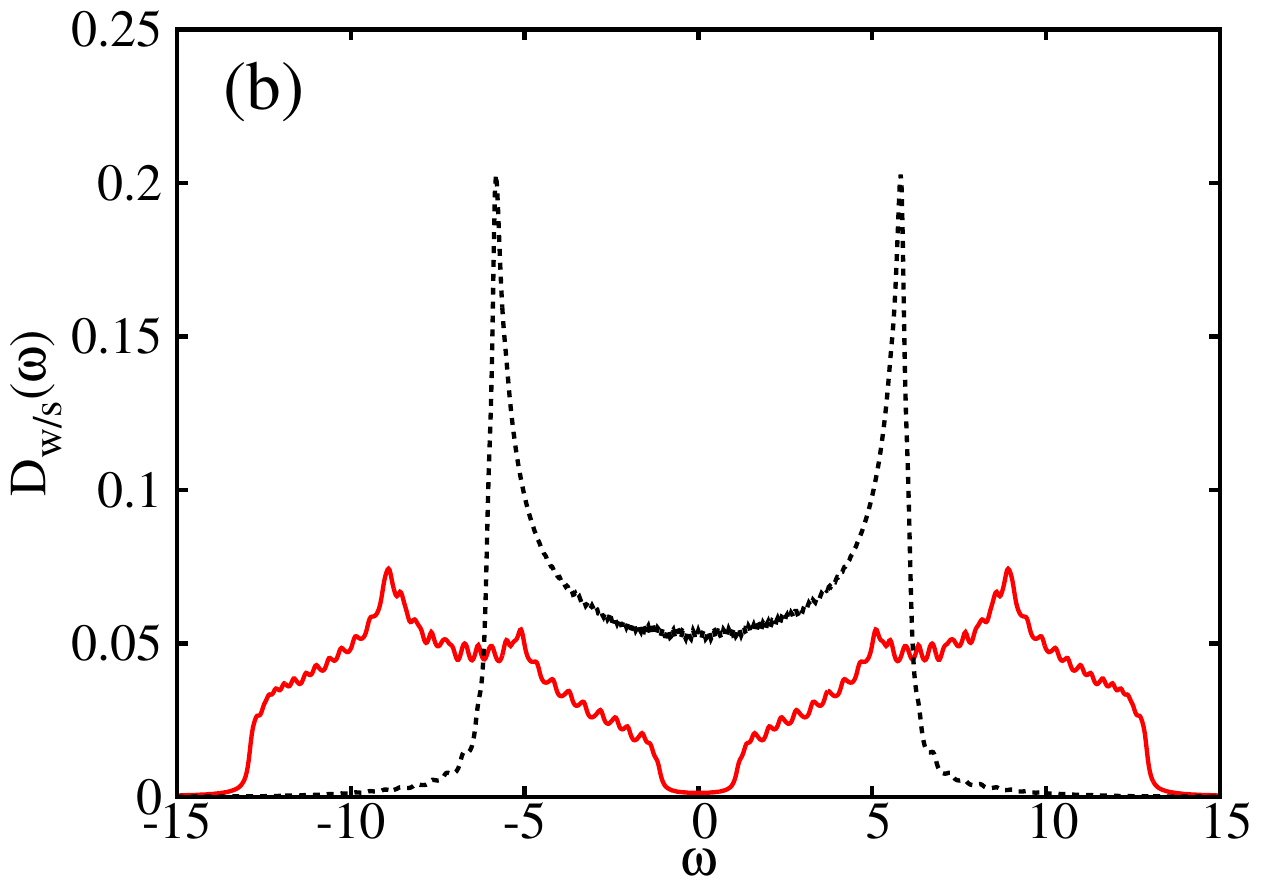}
\caption{\label{fig5} (Color online)
As in Fig.~\ref{fig4} but for an effective ladder model with
$N_{\text{leg}}=51$ legs.
}
\end{figure}

\begin{figure}[t]
\hspace*{7mm}
\includegraphics[width=0.38\textwidth]{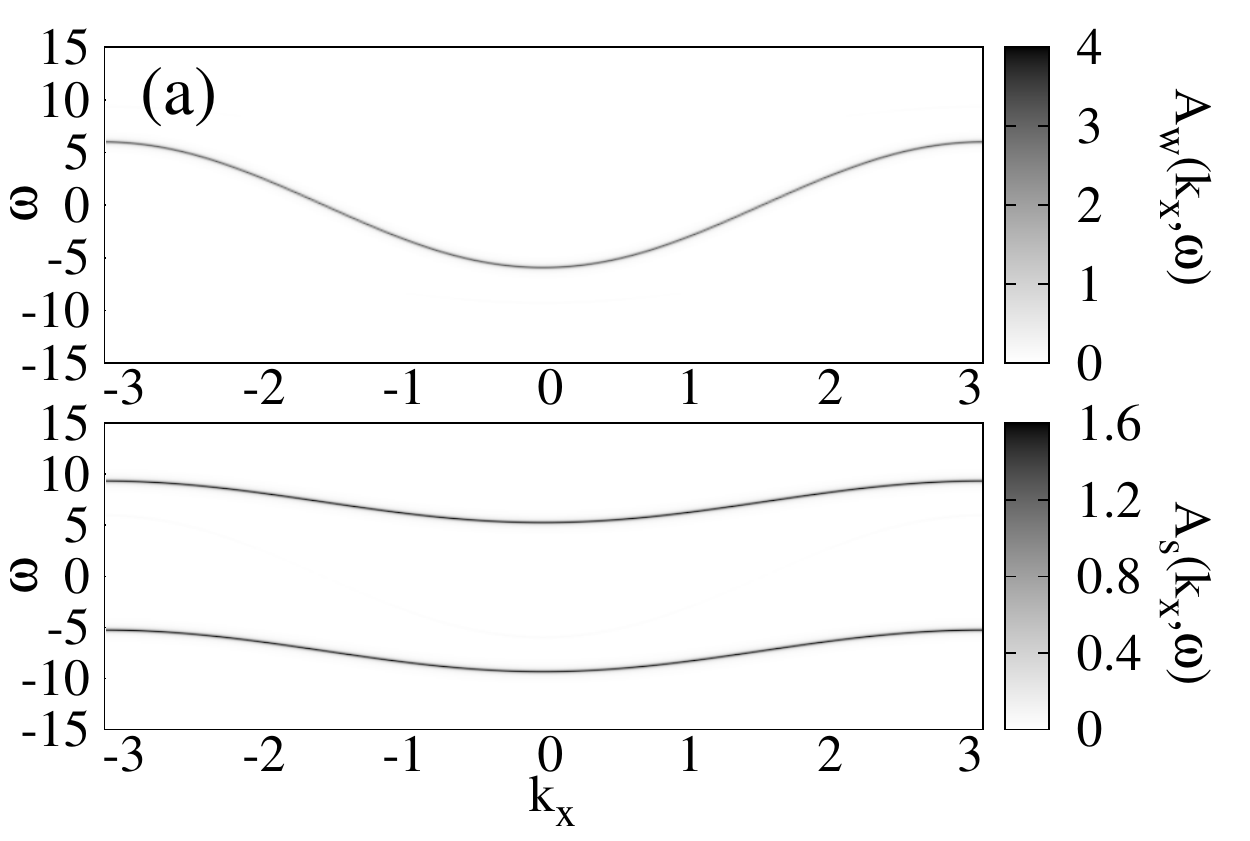}
\hspace*{-7.5mm}
\includegraphics[width=0.36\textwidth]{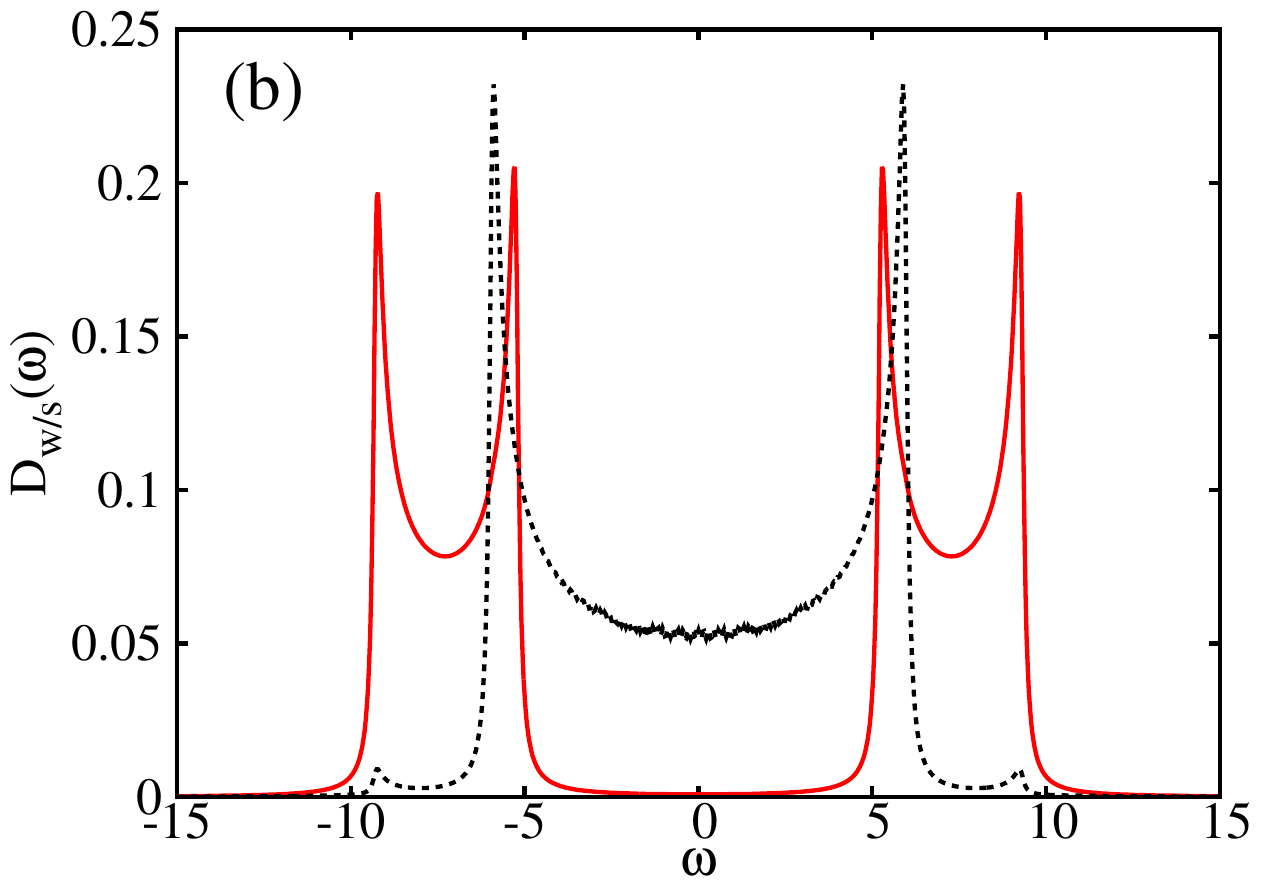}
\caption{\label{fig6} (Color online)
As in Fig.~\ref{fig4} but for a narrow ladder model
with $N_{\text{leg}}=3$ legs.
}
\end{figure}

The wire-substrate system can be mapped exactly to a 2D ladder-like system
with $N_{\text{imp}}=2L_yL_z+1=513$ legs (see Sec.~\ref{sec:mapping}).
Here, we examine NLMs with $N_{\text{leg}} < N_{\text{imp}}$. Figures~\ref{fig5} and~\ref{fig6} show results 
for the spectral functions and the DOS for $N_{\text{leg}}=51$ and $N_{\text{leg}}=3$, respectively. 
The other parameters are the same as for the case of the full system shown in Fig.~\ref{fig4}.
Overall, we find that the NLM can describe the full system correctly if two conditions are fulfilled.
First, the number of legs must be an odd number.
This condition can be easily interpreted: the NLM must include an even number of legs representing
the substrate (besides the leg representing the wire) to keep an equal number of degrees of freedom 
for the valence and conduction bands. Accordingly, 
an NLM must have at least three legs to describe a wire on an insulating substrate.
Second, the number of substrate legs must be large enough to represent the energy range of the
valence and conduction bands correctly, e.g., the band gap.
This condition can be achieved, however, using many fewer legs than in the full system,
i.e., for $N_{\text{leg}} \ll N_{\text{imp}}$.

For instance, the energy range of the substrate bands is already well
represented for $N_{\text{leg}}=51 \ll N_{\text{imp}}=513$ in Fig.~\ref{fig5}, although the spectral weight
distribution deviates visibly from that of the full system shown in Fig.~\ref{fig4}.
In contrast, Fig.~\ref{fig6} shows that the spectral functions and DOS
of the substrate are poorly represented by the NLM with $N_{\text{leg}}=3$ as
the spectral weight remains concentrated in two narrow bands.
In particular, the effective substrate band gap is $\Delta_{\text{s}}(N_{\text{leg}}=3) \approx 10$ 
and thus five times larger than the true gap $\Delta_{\text{s}} = 2$ while
the DOS exhibits van Hove singularities typical of two-leg ladders. 

Nevertheless, the wire spectral properties in the NLM with $N_{\text{leg}}\geq 3$ agree quantitatively 
with those of the 3D wire-substrate system close to the middle of the spectrum ($\omega \approx 0$).
Comparing with Fig.~\ref{fig4}(b) we see that the wire DOS in Fig.~\ref{fig5}(b) is reproduced correctly for
$N_{\text{leg}}=51$, while for $N_{\text{leg}}=3$ [Fig.~\ref{fig6}(b)] it is
substantially modified only where it overlaps with the substrate DOS ($\lvert \omega \rvert \agt 4$).
Therefore, our analysis suggests that
a three-leg NLM close to half-filling can be used to describe the low-energy
properties of a weakly interacting 
wire on an insulating substrate, at least qualitatively, as the degrees of freedom close to the Fermi energy 
($\epsilon_{\text{F}} \approx 0$) are 
correctly represented.
For instance, one could study the stability of Luttinger liquid features in a 1D conductor deposited
on an insulating substrate beyond the minimal models used previously~\cite{das01,das02,abd15}.  
However, for strong interactions or for a quantitative description of the
3D wire-substrate system by an effective NLM, we expect wider ladders and an
analysis of the convergence with $N_{\text{leg}}$ to be necessary.
Still, our results for noninteracting wires suggest that the required
 $N_{\text{leg}}$ can be significantly smaller than
the substrate size $N_{\text{imp}}$, making the NLMs amenable to standard methods
for quasi-1D correlated systems. The same holds true for interacting systems,
as demonstrated in Sec.~\ref{sec:QMC}. A detailed study of a correlated wire
with a Hubbard interaction can be found in~\cite{paper2}.

Finally, for a metallic substrate, we unsurprisingly find all single-electron eigenstates
to be delocalized over the entire wire-substrate system. 
(Of course there exist localized single-particle states in
the wire for energies above or below the metallic band, but theses states are not relevant for real
materials.)
Therefore, we do not observe any 1D features in a noninteracting wire coupled to a metallic substrate, but
this could be modified by interactions.
More decisively, we find that the spectral properties at the Fermi energy vary strongly with 
the number of legs in the NLM~(\ref{eq:ladder-hamiltonian}) even when $N_{\text{leg}}$ is as large as $51$. 
Therefore, we conclude that the full wire-substrate system cannot be represented even qualitatively by a quasi-1D
NLM if the substrate is metallic.

\subsection{\label{sec:QMC}Interacting wire}

Having established the usefulness of NLMs in the
noninteracting case, we briefly consider  a correlated wire with a repulsive
Hubbard interaction $U \geq 0$ [cf. Eq.~(\ref{eq:real-hubbard})]. A detailed
investigation of this problem, including the
Mott-insulating and Luttinger-liquid phases, can be found in~\cite{paper2}
where we also discuss the physics of the 1D Hubbard model. Here, we
compare spectral functions for the 3D wire-substrate problem to those of the
NLM~(\ref{eq:ladder-hamiltonian}) with $N_{\text{leg}}=3$.  The onsite
potential in the wire is $\epsilon_{\text w}=-U/2$, so that the
Hubbard bands are situated symmetrically around the middle of the substrate
band gap. The other parameters are taken to be the same as in
Figs.~\ref{fig4}--\ref{fig6}. We focus on a single set of parameters in the
metallic (Luttinger liquid) phase that exists away from half-filling. 
The interacting problem is solved by the continuous-time interaction
expansion (CT-INT) quantum Monte Carlo method~\cite{rubt05}, which can be
used to simulate the 3D wire-substrate system and NLMs with the same
numerical effort. For details see~\cite{paper2}. 

At half-filling, the full wire-substrate model contains $N_p=N_{\text{imp}}L_x$
electrons, whereas $N_p=N_{\text{leg}}L_x$ for the NLM. If the system is doped away from half-filling with a finite bulk doping $y \in (-1,1)$
[i.e., $N_p = (1 + y)N_{\text{imp}}L_x$ electrons in the full wire-substrate system or
$N_p=(1+y)N_{\text{leg}}L_x$ in the NLM], the chemical potential will lie in
one of the substrate bands.  This situation corresponds to a metallic substrate,
which is neither relevant for atomic wires  deposited on semiconducting substrates
nor expected to be well represented by an NLM. A more interesting and relevant
case is that of a finite wire doping $y_{\text w} \in (-1,1)$
($N_p = N_{\text{imp}}L_x + y_{\text w}L_x$ for the full wire-substrate model,
or $N_p=N_{\text{leg}}L_x + y_{\text w}L_x$ for the NLM) but a vanishing bulk doping
[$y\approx 0$ and $N_p/(N_{\text{imp}}L_x)\approx 1$]. In the latter, our
wire-substrate model can describe a quasi-1D conductor embedded in an
insulating 3D bulk system, as relevant for metallic wires on semiconducting substrates.
Besides In/Si(111)~\cite{sni10} above its critical temperature as well as
the Au/Ge(100)~\cite{blu11,blum12,naka12,park14,jong16}
and Bi/InSb(100)~\cite{ohts15} systems mentioned in Sec.~\ref{sec:intro},
Pt/Ge(100)~\cite{yaji13,yaji16}, Pb/Si(557)~\cite{Tegenkamp2005},
and dysprosium silicide nanowires on Si(001) surfaces~\cite{Wanke2011}
are also known to be metallic.

\begin{figure}
\includegraphics[width=0.5\textwidth]{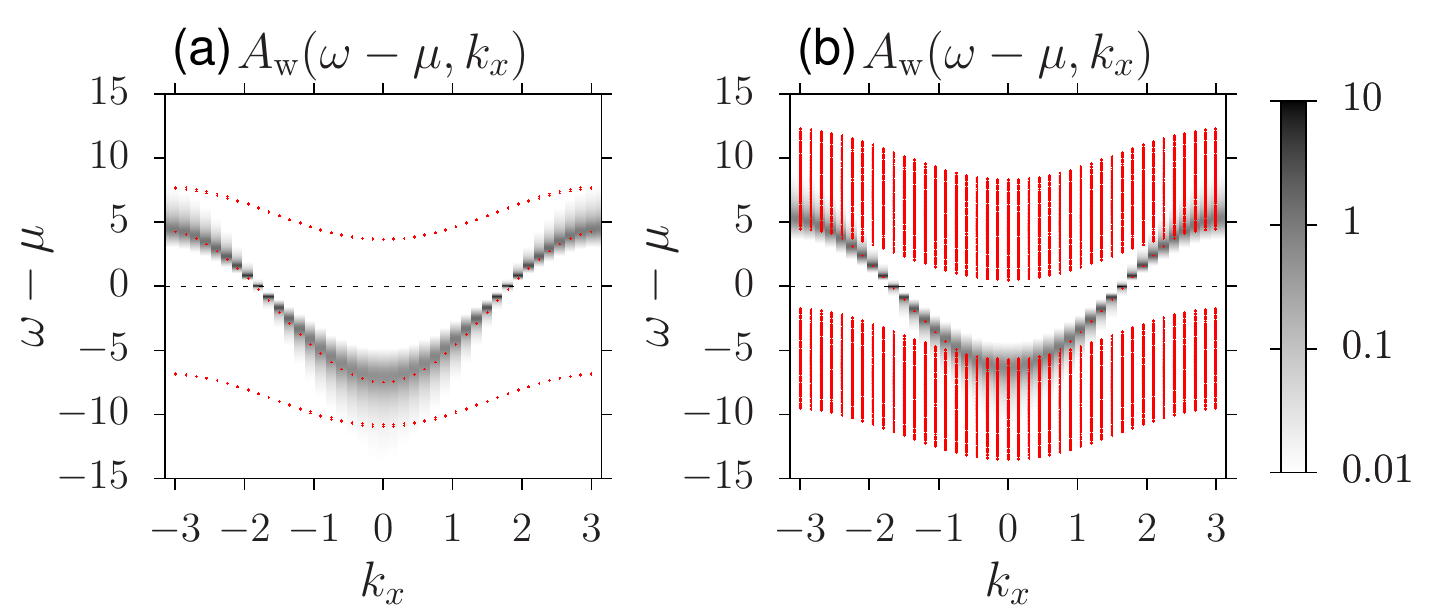}
\includegraphics[width=0.5\textwidth]{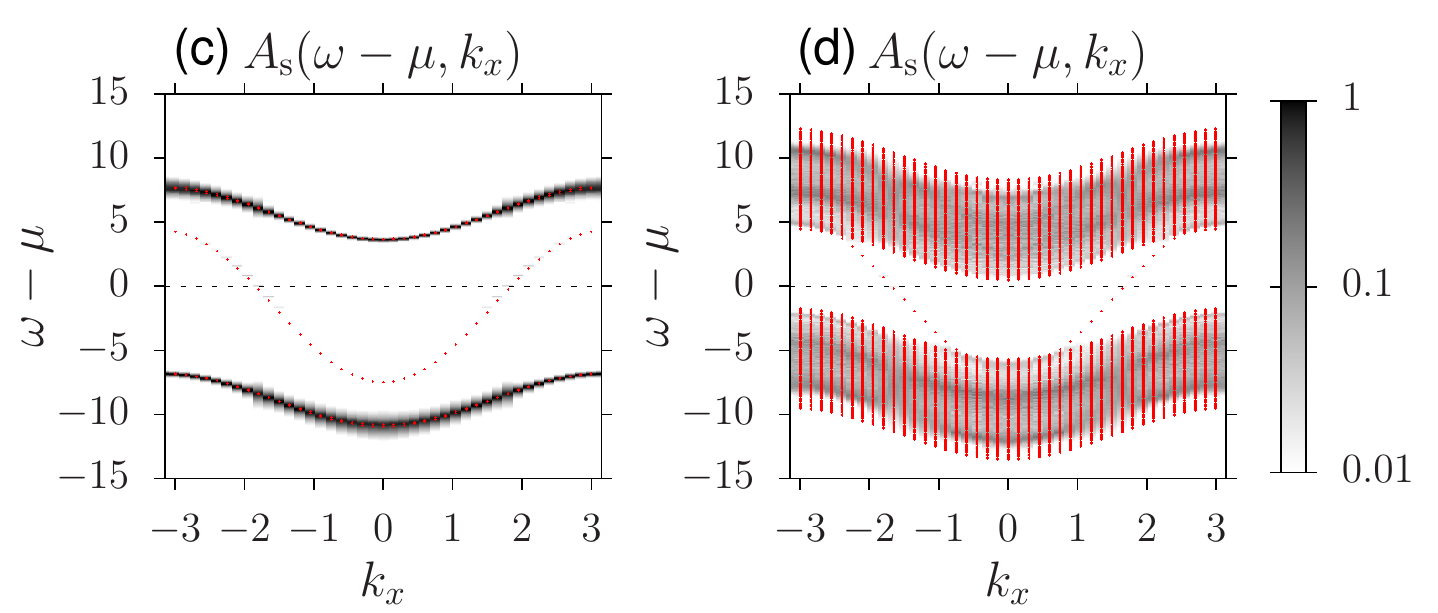}
\caption{\label{fig:qmc}
(Color online)
Spectral functions calculated for an interacting wire ($U=4$) as obtained from CT-INT QMC simulations with $\beta=15$.
The chemical potential was tuned to obtain a wire doping $y_{\text{w}}\approx 12.5\%$. 
%of about 5.25 electrons.
Top: wire spectral function $A_{\text{w}}(\omega,k_x)$ for (a) the three-leg
NLM and (b) the 3D wire-substrate model. Bottom: (c) substrate spectral
function $A_{\text{s}}(\omega,k_x)$ for the three-leg NLM, (d) substrate
spectral function $A_{\text{s}}(\omega,k_x)$ of the full model averaged over
the chains at the minimal and the maximal distance from the wire. In all
cases, $L_x=42$, and for the 3D substrate lattice $L_y=42$ and $L_z=10$. 
All other parameters as in Figs.~\ref{fig4}--\ref{fig6}. Red symbols illustrate the noninteracting energy levels.}
\end{figure}

We calculated the finite-temperature analogs of the single-particle spectral functions 
$A_{\text{w}}(\omega,k_x)$, $A(\omega,k_x,n)$, and $A_{\text{s}}(\omega,k_x,y,z)$
defined at the beginning of Sec.~\ref{sec:ladder}. For example, the wire
spectral function is given by
\begin{align}\label{eq:akw}\nonumber
  A_{\text{w}}(\omega,k_x) 
  &=
  \frac{1}{Z}\sum_{ij}
  {|\langle {i}| d_{\text{w}k_x\sigma} |{j}\rangle|}^2 (e^{-\beta E_i}+e^{-\beta E_j})
  \\
  &\hspace*{4em}\times
  \delta(\Delta_{ji}-\omega)
\end{align}
and can be obtained from the single-particle Green function
$G(k,\tau)=\langle d^{\dag}_{\text{w} k \sigma} (\tau) d^{\phantom{\dag}}_{\text{w} k \sigma}(0) \rangle$
by analytic continuation \cite{bea04}. In Eq.~(\ref{eq:akw}), $Z$ is the
grand-canonical partition function, $|{i}\rangle$ is an eigenstate with
energy $E_i$, and $\Delta_{ji}=E_j-E_i$. Similarly, we can obtain $A(\omega,k_x,n)$ in the NLM 
and $A_{\text{s}}(\omega,k_x,y,z)$ for specific chains in the substrate of the 3D wire-substrate model. 
Because a full substrate average is expensive with the CT-INT method,
substrate properties were averaged over the chains at the minimal 
($y=y_0,z=1$) and maximal ($y=y_0+L_y/2,z=L_z$) distance from the wire.

For sufficiently weak coupling $U$, we find that a metallic wire is realized
in both the 3D wire-substrate model and in the NLM at low wire doping.
As an example, Fig.~\ref{fig:qmc} compares the spectral functions
$A_{\text w}(\omega,k_x)$ and $A_{\text s}(\omega,k_x,y=y_0,z=1)$
of the 3D wire-substrate model with the spectral functions
$A_{\text w}(\omega,k_x)$ and $A_{\text s}(\omega,k_x)$
of the three-leg NLM at $U=4$. The chemical potential was set to 
$\mu = 1.58$ for the NLM and to $\mu=0.60$ for the full model, 
corresponding to a total doping of 5.25(1) electrons (or $y_{\text w}\approx 12.5\%$ for $L_x=42$).
A comparison of Figs.~\ref{fig:qmc}(a) and~(b)
reveals that the wire spectral functions  of the
two models agree close to the Fermi energy, and that there exist gapless excitations
predominantly localized in the wire. In contrast, the substrate spectral functions in
Figs.~\ref{fig:qmc}(c) and~(d) differ significantly, as already observed for
the noninteracting case. Nevertheless, both models exhibit a vanishingly small
weight at the Fermi energy.

These QMC results confirm that---similar to the noninteracting case---the
low-energy excitations of a metallic interacting wire in the 3D
wire-substrate model are well represented, at least qualitatively, by the
three-leg NLM for moderate couplings $U$. For strong interactions, however,
the low-energy excitations can be delocalized in the substrate and the
NLM approximation becomes less accurate. This regime is discussed in
detail in~\cite{paper2}.

\section{\label{sec:Conclusion}Conclusions}

We introduced lattice models for correlated atomic wires on noninteracting
metallic or insulating substrates, and showed that they can be mapped exactly
onto ladder-like 2D lattices. 
The first leg corresponds to the atomic wire, while the other legs represent
successive shells of substrate sites with increasing distance from the wire. 
We investigated the approximation of narrow ladder models 
that take only a few shells around the wire into account. 
Our results suggest that the low-energy physics of (weakly-interacting) 
wires on insulating substrates can be
described by ladder models with three or more legs. 
These models can be studied with well-established methods for 1D correlated systems 
such as the DMRG~\cite{whi92,whi93,sch05,jec08a}
or field-theoretical techniques (e.g., bosonization and the  
renormalization
group)~\cite{sol79,giamarchi07,Schoenhammer,Gogolin,Tsvelik}. We believe that 
the approach developed here can shed new light on the quasi-1D physics and correlation effects 
to be found in atomic wires deposited on semiconducting substrates. 
As a first application, we investigate
Mott and Luttinger liquid phases of a Hubbard-type wire on an
insulating substrate using DMRG and QMC methods in~\cite{paper2}.

In the future, we plan to apply this approach to real metallic wire-substrate systems.
The wire-substrate model defined in Sec.~\ref{sec:model} can easily be generalized to
achieve a more realistic description of specific experiments. In particular,
one can use first-principles band structures and
hybridizations~\cite{Cano07,abinitio2,abinitio3,abinitio4} for the
noninteracting part of the Hamiltonian. However, determining the interaction
parameters from first-principles simulations remains an open problem~\cite{Lee2011}.

\begin{acknowledgments}
This work was supported by the German Research Foundation (DFG) through SFB
1170 ToCoTronics and the Research Unit \textit{Metallic nanowires on the atomic scale: Electronic
and vibrational coupling in real world systems} (FOR~1700, grant No.~JE~261/1-1).
The authors gratefully acknowledge the computing time granted by the John
von Neumann Institute for Computing (NIC) and provided on the supercomputer
JURECA \cite{Juelich} at the J\"{u}lich Supercomputing Centre.
\end{acknowledgments}

  \end{document}